\documentclass[12pt]{book}

\usepackage[numbers]{natbib}
\bibpunct{[}{]}{;}{n}{,}{,}
\usepackage{amsmath,amssymb,amsmath,amsthm}
\usepackage{fullpage}
\usepackage{graphicx,subfigure}
\usepackage[chapter]{algorithm}
\usepackage{algorithmic}
\usepackage{bm}
\usepackage{url}
\usepackage[pdftex,bookmarks,colorlinks=true,citecolor=blue]{hyperref}
\usepackage{listings}
\usepackage{xcolor}
\usepackage{color}

\newtheorem{theorem}{Theorem}[chapter]

\newtheorem{fact}[theorem]{Fact}

\newtheorem{property}{Property}[chapter]

\def\A{{\bf A}}
\def\a{{\bf a}}
\def\B{{\bf B}}
\def\bb{{\bf b}}
\def\C{{\bf C}}
\def\c{{\bf c}}
\def\D{{\bf D}}

\def\g{{\bf g}}

\def\G{{\bf G}}
\def\H{{\bf H}}
\def\I{{\bf I}}
\def\K{{\bf K}}
\def\LL{{\bf L}}

\def\PP{{\bf P}}
\def\Q{{\bf Q}}

\def\R{{\bf R}}
\def\rr{{\bf r}}
\def\S{{\bf S}}

\def\T{{\bf T}}

\def\U{{\bf U}}
\def\u{{\bf u}}
\def\V{{\bf V}}
\def\v{{\bf v}}
\def\W{{\bf W}}
\def\w{{\bf w}}
\def\X{{\bf X}}
\def\x{{\bf x}}
\def\Y{{\bf Y}}
\def\y{{\bf y}}
\def\Z{{\bf Z}}
\def\z{{\bf z}}
\def\0{{\bf 0}}
\def\1{{\bf 1}}

\def\NM{{\mathcal N}}
\def\OM{{\mathcal O}}
\def\PM{{\mathcal P}}
\def\SM{{\mathcal S}}

\def\RB{{\mathbb R}}

\def\EB{{\mathbb E}}

\def\Si{\mbox{\boldmath$\Sigma$\unboldmath}}

\def\Lam{\mbox{\boldmath$\Lambda$\unboldmath}}

\def\argmin{\mathop{\rm argmin}}

\def\nnz{\mathrm{nnz}}
\def\poly{\mathrm{poly}}

\def\rk{\mathrm{rank}}

\def\st{\mathsf{s.t.}}

\def\etal{{\em et al.\/}\,}

\newcommand{\red}[1]{{\color{red}#1}}

\newcommand{\comment}[1]{}

\begin{document}

\title{\bf A Practical Guide to Randomized Matrix Computations with MATLAB Implementations\thanks{Sample MATLAB code with demos is available at https://github.com/wangshusen/RandMatrixMatlab.}}
\author{
Shusen Wang\\
wssatzju@gmail.com
%\vspace{10cm}
}
%\date{November 2015}

\maketitle
\cleardoublepage \pagenumbering{roman}
\tableofcontents
\clearpage
\setcounter{page}{0}
\pagenumbering{arabic}

\chapter*{Abstract}
\addcontentsline{toc}{chapter}{Abstract}%

Matrix operations such as matrix inversion, eigenvalue decomposition, singular value decomposition are ubiquitous in real-world applications. Unfortunately, many of these matrix operations so time and memory expensive that they are prohibitive when the scale of data is large. In real-world applications, since the data themselves are noisy, machine-precision matrix operations are not necessary at all, and one can sacrifice a reasonable amount of accuracy for computational efficiency.

In recent years, a bunch of randomized algorithms have been devised to make matrix computations more scalable. Mahoney \cite{mahoney2011ramdomized} and Woodruff \cite{woodruff2014sketching} have written excellent but very technical reviews of the randomized algorithms. Differently, the focus of this paper is on intuition, algorithm derivation, and implementation. This paper should be accessible to people with knowledge in elementary matrix algebra but unfamiliar with randomized matrix computations. The algorithms introduced in this paper are all summarized in a user-friendly way, and they can be implemented in lines of MATLAB code. The readers can easily follow the implementations even if they do not understand the maths and algorithms.

\vspace{5mm}

\noindent
{\bf Keywords:}
matrix computation, randomized algorithms, matrix sketching, random projection, random selection,
least squares regression, randomized SVD, matrix inversion, eigenvalue decomposition, kernel approximation, the Nystr\"om method.

%%%%%%%%%%%%%%%%%%%%%%%%%%%%%%%%%%%%%%%%%%%%%%%%%%%%%%%%%%%%%%%%%%%%%%%%%%%%%%
%%%%%%%%%%%%%%%%%%%%%%%%%%%%%%%%%%%%%%%%%%%%%%%%%%%%%%%%%%%%%%%%%%%%%%%%%%%%%%
%%%%%%%%%%%%%%%%%%%%%%%%%%%%%%%%%%%%%%%%%%%%%%%%%%%%%%%%%%%%%%%%%%%%%%%%%%%%%%
%%%%%%%%%%%%%%%%%%%%%%%%%%%%%%%%%%%%%%%%%%%%%%%%%%%%%%%%%%%%%%%%%%%%%%%%%%%%%%

\chapter{Introduction} \label{sec:introduction}

Matrix computation plays a key role in modern data science.
However, matrix computations such as matrix inversion, eigenvalue decomposition, SVD, etc,
are very time and memory expensive,
which limits their scalability and applications.
To make large-scale matrix computation possible,
randomized matrix approximation techniques have been proposed and widely applied.
Especially in the past decade, remarkable progresses in randomized numerical linear algebra has been made,
and now large-scale matrix computations are no longer impossible tasks.

This paper reviews the most recent progresses of randomized matrix computation.
The papers written by Mahoney \cite{mahoney2011ramdomized} and Woodruff \cite{woodruff2014sketching}
provide comprehensive and rigorous reviews of the randomized matrix computation algorithms.
However, their focus are on the theoretical properties and error analysis techniques,
and readers unfamiliar with randomized numerical linear algebra can have difficulty when implementing their algorithms.

Differently, the focus of this paper is on intuitions and implementations,
and the target readers are those who are familiar with basic matrix algebra but has little knowledge in randomized matrix computations.
All the algorithms in this paper are described in a user-friend way.
This paper also provides MATLAB implementations of the important algorithms.
MATLAB code is easy to understand\footnote{If your are unfamiliar with a MATLAB function,
you can simply type ``$\mathrm{help+functionname}$'' in MATLAB and read the documentation.},
easy to debug, and easy to translate to other languages.
The users can even directly use the provided MATLAB code without understanding it.

This paper covers the following topics:
\begin{itemize}
\item
    Chapter~\ref{sec:elementary} briefly reviews some matrix algebra preliminaries.
    This chapter can be skipped if the reader is familiar with matrix algebra.
\item
    Chapter~\ref{sec:sketching} introduces the techniques for generating a sketch of a large-scale matrix.
\item
    Chapter~\ref{sec:lsr} studies the least squares regression (LSR) problem where $n \gg d$.
\item
    Chapter~\ref{sec:ksvd} studies efficient algorithms for computing the $k$-SVD of arbitrary matrices.
\item
    Chapter~\ref{sec:spsd} introduces techniques for sketching symmetric positive semi-definite (SPSD) matrices.
    The applications includes spectral clustering, kernel methods (e.g.\ Gaussian process regression and kernel PCA),
    and second-order optimization (e.g.\ Newton's method).
\end{itemize}

%%%%%%%%%%%%%%%%%%%%%%%%%%%%%%%%%%%%%%%%%%%%%%%%%%%%%%%%%%%%%%%%%%%%%%%%%%%%%%
%%%%%%%%%%%%%%%%%%%%%%%%%%%%%%%%%%%%%%%%%%%%%%%%%%%%%%%%%%%%%%%%%%%%%%%%%%%%%%
%%%%%%%%%%%%%%%%%%%%%%%%%%%%%%%%%%%%%%%%%%%%%%%%%%%%%%%%%%%%%%%%%%%%%%%%%%%%%%
%%%%%%%%%%%%%%%%%%%%%%%%%%%%%%%%%%%%%%%%%%%%%%%%%%%%%%%%%%%%%%%%%%%%%%%%%%%%%%

\chapter{Elementary Matrix Algebra} \label{sec:elementary}

This chapter defines the matrix notation and goes through the very basics of matrix decompositions.
Particularly, the singular value decomposition (SVD), the QR decomposition, and the Moore-Penrose inverse are used throughout this paper.

\section{Notation}

Let $\A = [a_{i j}]$ be a matrix,
$\a = [a_i]$ be a column vector,
and $a$ be a scalar.
The $i$-th row and $j$-th column of $\A$ are denoted by $\a_{i:}$ and $\a_{:j}$, respectively.
When there is no ambiguity, either column or row can be written as $\a_l$.
Let $\I_n$ be the $n\times n$ identity matrix, that is, the diagonal entries are ones and off-diagonal entries are zeros.
The column space (the space spanned by the columns) of $\A$ is the set of all possible linear combinations of its column vectors.
Let $[n]$ be the set $\{ 1, 2, \cdots , n\}$.
Let $\nnz (\A)$ be the number of nonzero entries of $\A$.

The squared vector $\ell_2$ norm is defined by
\[
\| \a\|_2^2 \; = \; {\sum_i a_i^2}.
\]
The squared matrix Frobenius norm is defined by
\[
\| \A \|_F \; = \; {\sum_{i j} a_{i j}^2} ,
\]
and the matrix spectral norm is defined by
\[
\| \A \|_2 \; = \; \max_{\x \neq \0} \frac{\|\A \x\|_2}{ \| \x\|_2} .
\]

\section{Matrix Decompositions}

{\bf QR decomposition.}
Let $\A$ be an $m\times n$ matrix with $m \geq n$.
The QR decomposition of $\A$ is
\[
\A \; = \; \underbrace{\Q_\A}_{m\times n} \underbrace{\R_\A}_{n\times n}.
\]
The matrix $\Q_\A$ has orthonormal columns, that is,
$\Q_{\A}^T \Q_{\A} = \I_n$.
The matrix $\R_\A$ is upper triangular, that is, for all $i < j$, the $(i,j)$-th entry of $\R_\A$ is zero.

{\bf SVD.}
Let $\A$ be an $m\times n$ matrix and $\rho = \rk (\A)$.
The condensed singular value decomposition (SVD) of $\A$ is
\[
\underbrace{\A }_{m\times n}
\; = \; \underbrace{\U_\A}_{m\times \rho} \underbrace{\Si_\A}_{\rho \times \rho} \underbrace{\V_\A^T}_{\rho\times n}
\; = \; \sum_{i = 1}^{\rho} \sigma_{\A,i} \u_{\A,i} \v_{\A , i}^T.
\]
Here $\sigma_{\A,1} \geq \cdots \geq \sigma_{\A,\rho} > 0$ are the singular values,
$\u_{\A,1} , \cdots , \u_{\A,\rho} \in \RB^m$ are the left singular vectors,
and $\v_{\A,1} , \cdots , \v_{\A,\rho} \in \RB^n$ are the right singular vectors.
Unless otherwise specified, ``SVD'' refers to the condensed SVD.

{\bf $k$-SVD.}
In applications such as the principal component analysis (PCA), latent semantic indexing (LSI), word2vec, spectral clustering,
we are only interested in the top $k$ ($\ll m, n$) singular values and singular vectors.
The rank $k$ truncated SVD ($k$-SVD) is denoted by
\[
\A_k
\; := \;
\sum_{i = 1}^{k} \sigma_{\A,i} \u_{\A,i} \v_{\A , i}^T
\; = \; \underbrace{\U_{\A,k}}_{m\times k} \underbrace{\Si_{\A,k}}_{k\times k} \underbrace{\V_{\A,k}^T}_{k\times n} .
\]
Here $\U_{\A, k}$ consists of the first $k$ singular vectors of $\U_\A$, and $\Si_{\A,k}$ and $\V_{\V,k}$ are analogously defined.
Among all the $m\times n$ rank $k$ matrices,
$\A_k$ is the closest approximation to $\A$ in that
\[
\A_k \; = \;
\argmin_{\X} \|\A - \X\|_F^2
\; = \;
\argmin_{\X} \|\A - \X\|_2^2 ,
\qquad
\st \; \rk (\X) \leq k.
\]

{\bf Eigenvalue decomposition.}
The eigenvalue decomposition of an $n\times n$ symmetric matrix $\A$ is defined by
\[
\A
\; = \; \U_\A \Lam_\A \U_\A^T
\; = \; \sum_{i=1}^n \lambda_{\A,i} \u_{\A , i} \u_{\A , i}^T .
\]
Here $\lambda_{\A,1} \geq \cdots \geq \lambda_{\A,n}$ are the eigenvalues of $\A$,
and $\u_{\A,1} , \cdots , \u_{\A,n} \in \RB^n$ are the corresponding eigenvectors.
A symmetric matrix $\A$ is called symmetric positive semidefinite (SPSD) if and only if all the eigenvalues are nonnegative.
If $\A$ is SPSD,
its SVD and eigenvalue decomposition are identical.

\section{Matrix (Pseudo) Inverse and Orthogonal Projector}

For an $n\times n$ square matrix $\A$,
the matrix inverse exists if $\A$ is non-singular ($\rk (\A) = n$).
Let $\A^{-1}$ be the inverse of $\A$.
Then $\A \A^{-1} = \A^{-1} \A = \I_n$.

Only square and full rank matrices have inverse.
For the general rectangular matrices or rank deficient matrices,
matrix pseudo-inverse is used as a generalization of matrix inverse.
The book \cite{adi2003inverse} offers a comprehensive study of the pseudo-inverses.

The Moore-Penrose inverse is the most widely used pseudo-inverse, which is defined by
\[
\A^\dag \; := \;
\V_{\A} \Si_\A^{-1} \U_{\A}^T .
\]
Let $\A$ be any $m\times n$ and rank $\rho$ matrix.
Then
\[
\A \A^\dag \; = \; \U_{\A} \Si_\A \underbrace{\V_{\A}^T  \V_{\A}}_{=\I_\rho} \Si_\A^{-1} \U_{\A}^T
\; = \; \underbrace{\U_{\A}}_{m\times \rho} \underbrace{\U_{\A}^T}_{\rho \times m} ,
\]
which is a orthogonal projector.
It is because for any matrix $\B$,
the matrix $\A \A^\dag \B = \U_\A \U_\A^T \B $ is the projection of $\B$ onto the column space of $\A$.

%and
%\[
%\A^\dag \A \; = \;  \V_{\A} \Si_\A^{-1} \underbrace{\U_{\A}^T \U_{\A}}_{=\I_\rho} \Si_\A \V_{\A}^T
%\; = \; \underbrace{\V_{\A}}_{n\times \rho} \underbrace{\V_{\A}^T}_{\rho \times n}.
%\]

\section{Time and Memory Costs}

{\bf The time complexities} of the matrix operations are listed in the following.
\begin{itemize}
\item
    Multiplying an $m\times n$ matrix $\A$ by an $n\times p$ matrix $\B$: $\OM (mnp)$ float point operations (flops) in general,
    and $\OM (p \cdot \nnz (\A))$ if $\A$ is sparse. Here $\nnz (\A)$ is the number of nonzero entries of $\A$.
\item
    QR decomposition, SVD, or Moore-Penrose inverse of an $m\times n$ matrix ($m \geq n$): $\OM (m n^2)$ flops.
\item
    $k$-SVD of an $m\times n$ matrix: $\OM (n m k)$ flops
    (assuming that the spectral gap and the logarithm of error tolerance are constant)
\item
    Matrix inversion or full eigenvalue decomposition of an $n\times n$ matrix: $\OM (n^3)$ flops
\item
    $k$-eigenvalue decomposition of an $n\times n$ matrix: $\OM (n^2 k)$ flops.
\end{itemize}

{\bf Pass-efficient} means that the algorithm goes constant passes through the data.
For example, the Frobenius norm of a matrix can be computed pass-efficiently,
because each entry is visited only once.
In comparison, the spectral norm cannot be computed pass-efficiently,
because the algorithm goes at least $\log \frac{1}{\epsilon}$ passes through the matrix, which is not constant.
Here $\epsilon$ indicates the desired precision.

{\bf Memory cost.}
If an algorithm scans a matrix for constant passes,
the matrix can be placed in large volume disks, so the memory cost is not a bottleneck.
However, if an algorithm goes through a matrix for many passes (not constant passes),
the matrix should be placed in memory, otherwise the swaps between memory and disk would be highly expensive.
In this paper, memory cost means the number of entries frequently visited by the algorithm.

%%%%%%%%%%%%%%%%%%%%%%%%%%%%%%%%%%%%%%%%%%%%%%%%%%%%%%%%%%%%%%%%%%%%%%%%%%%%%%
%%%%%%%%%%%%%%%%%%%%%%%%%%%%%%%%%%%%%%%%%%%%%%%%%%%%%%%%%%%%%%%%%%%%%%%%%%%%%%
%%%%%%%%%%%%%%%%%%%%%%%%%%%%%%%%%%%%%%%%%%%%%%%%%%%%%%%%%%%%%%%%%%%%%%%%%%%%%%
%%%%%%%%%%%%%%%%%%%%%%%%%%%%%%%%%%%%%%%%%%%%%%%%%%%%%%%%%%%%%%%%%%%%%%%%%%%%%%

\chapter{Matrix Sketching} \label{sec:sketching}

Let $\A \in \RB^{m\times n}$ be the given matrix,
$\S \in \RB^{n\times s}$ be a sketching matrix, e.g.\ random projection or column selection matrix,
and $\C  = \A \S \in \RB^{m\times s}$ be a sketch of $\A$.
The size of $\C$ is much smaller than $\A$, but $\C$ preserves some important properties of $\A$.

\section{Theoretical Properties} \label{sec:sketching_theory}

The sketching matrix is useful if it has either or both of the following properties.
The two properties are important, and the readers should try to understand them.

\begin{property}[Subspace Embedding]
For a fixed $m\times n$ ($m \ll n$) matrix $\A$ and all $m$-dimension vector $\y$, the inequality
\[
\frac{1}{\gamma} \; \leq \; \frac{\| \y^T \A \S \|_2^2}{ \|\y^T \A \|_2^2} \; \leq \; \gamma
\]
holds with high probability.
Here $\S \in \RB^{n\times s}$ ($s \ll n$) is a certain sketching matrix.
\end{property}

The subspace embedding property can be intuitively understood in the following way.
For all $n$ dimensional vectors $\x$ in the row
space of $\A$ (a rank $m$ subspace within $\RB^n$),\footnote{Thus there always exists an $m$ dimensional vector $y$ such that $\x$ can be expressed as $\x = \y^T \A$.}
the length of vector $\x$ does not change much after sketching: $\|\x\|_2^2 \approx \|\x\S\|_2^2$.
This property can be applied to speedup the $\ell_2$ regression problems.

\begin{property}[Low-Rank Approximation]
Let $\A$ be any $m\times n$ matrix and $k$ be any positive integer far smaller than $m$ and $n$.
Let $\C = \A \S \in \RB^{m\times s}$ where $\S \in \RB^{n\times s}$ is a certain sketching matrix and $s \geq k$.
The Frobenius norm error bound\footnote{Spectral norm bounds should be more interesting.
However, spectral norm error is difficult to analyze, and existing spectral norm bounds are ``weak'' for their factors $\eta$ are far greater than 1.}
\[
\| \A - \C \C^\dag \A \|_F^2 \; \leq \; \eta \| \A - \A_k\|_F^2
\]
holds with high probability for some $\eta \geq 1$.

The following error bound is stronger and more interesting:
\[
\min_{\rk(\X) \leq k} \, \| \A - \C \X \|_F^2 \; \leq \; \eta \| \A - \A_k\|_F^2 .
\]
It is stronger because $\| \A - \C \C^\dag \A \|_F^2 \leq \min_{\rk(\X) \leq k} \, \| \A - \C \X \|_F^2$.
\end{property}

Intuitively speaking, the low-rank approximation property means that
the columns of $\A_k$ are almost in the column space of $\C = \A \PP$.
The low-rank approximation property enables us to solve $k$-SVD more efficiently (for $k\leq s$).
Later on we will see that computing the $k$-SVD of $\C \C^\dag \A$ is less expensive than the $k$-SVD of $\A$.

The two properties can be verified by a few lines of MATLAB code.
The readers are encouraged to have a try.
With a proper sketching method and a relatively large $s$, both $\gamma$ and $\eta$ should be near one.

\section{Random Projection} \label{sec:randprojection}

The section presents three matrix sketching techniques:
Gaussian projection, subsampled randomized Hadamard transform (SRHT), and count sketch.
Gaussian projection and SRHT can be combined with count sketch.

\subsection{Gaussian Projection}

The ${n\times s} $ Gaussian random projection matrix $\S$ is a matrix is formed by
$\S  = \frac{1}{\sqrt{s}} \G$, where each entry of $\G$ is sampled i.i.d.\ from $\NM (0, 1)$.
The Gaussian projection is also well knows as the Johnson-Lindenstrauss transform due to the seminal work \cite{johnson1984extensions}.
Gaussian projection can be implemented in four lines of MATLAB code.
\vspace{3mm}
\begin{lstlisting}
function [C] = GaussianProjection(A, s)
n = size(A, 2);
S = randn(n, s) / sqrt(s);
C = A * S;
\end{lstlisting}
\vspace{3mm}
Gaussian projection has the following properties:
\begin{itemize}
\item
    Time cost: $\OM (m n s)$
\item
    Theoretical guarantees
    \begin{enumerate}
    \item
        When $s = \OM ( m / \epsilon^{2} )$,
        the subspace embedding property with $\gamma = 1+\epsilon$ holds with high probability.
    \item
        When $s = \frac{k}{\epsilon} + 1$, the low-rank approximation property with $\eta = 1+\epsilon$ holds in expectation \cite{boutsidis2011near}.
    \end{enumerate}
\item
    Advantages
    \begin{enumerate}
    \item
        Easy to implement: four lines of MATLAB code
    \item
        $\C$ is a very high quality sketch of $\A$
    \end{enumerate}
    \item
        Disadvantages:
        \begin{enumerate}
        \item
            High time complexity to perform matrix multiplication
        \item
            Sparsity is destroyed: $\C$ is dense even if $\A$ is sparse
        \end{enumerate}
\end{itemize}

\subsection{Subsampled Randomized Hadamard Transform (SRHT)} \label{sec:sketching:srht}

The Subsampled Randomized Hadamard Transform (SRHT) matrix is defined by $\S = \frac{1}{\sqrt{s n}} \D \H_n \PP$, where
\begin{itemize}
\item
    $\D \in \RB^{n\times n}$ is a diagonal matrix with diagonal entries sampled uniformly from $\{+1, -1\}$;
\item
    $\H_n \in \RB^{n\times n}$ is defined recursively by
    \[
    \H_n = \left[
             \begin{array}{cc}
               \H_{n/2} & \H_{n/2} \\
               \H_{n/2} & -\H_{n/2} \\
             \end{array}
           \right]
    \qquad \textrm{ and } \qquad
    \H_2 = \left[
             \begin{array}{cc}
               +1 & +1 \\
               +1 & -1 \\
             \end{array}
           \right];
    \]
    For all $\y \in \RB^n$,
    the matrix vector product $ \y^T \H_n$ can be performed in $\OM (n \log n)$ time
    by the fast Walsh--Hadamard transform algorithm in a divide-and-conquer fashion;
\item
    $\PP \in \RB^{n\times s}$ samples $s$ from the $n$ columns.
\end{itemize}
SRHT can be implemented in nine lines of MATLAB code below.
Notice that this implementation of SRHT is has $\OM ( m N \log N)$
($N \geq n$ is a power of two)
time complexity, which is not efficient.
\vspace{3mm}
\begin{lstlisting}
function [C] = srht(A, s)
n = size(A, 2);
sgn = randi(2, [1, n]) * 2 - 3; % one half are +1 and the rest are -1
A = bsxfun(@times, A, sgn); % flip the signs of each column w.p. 50%
n = 2^(ceil(log2(n)));
C = (fwht(A', n))'; % fast Walsh-Hadarmard transform
idx = sort(randsample(n, s));
C = C(:, idx); % subsampling
C = C * (n / sqrt(s));
\end{lstlisting}
\vspace{3mm}
The SRHT matrix has the following properties:
\begin{itemize}
\item
    Time complexity:
    the matrix product $\A \S$ can be performed in $\OM (m n \log s)$ time,
    which makes SRHT more efficient than Gaussian projection.
    (Unfortunately, the MATLAB code above does not have such low time complexity.)
\item
    Theoretical property:
    when $s = \OM (\epsilon^{-2}  (m + \log n ) \log m)$, SRHT satisfies the subspace embedding property with $\gamma = 1+\epsilon$ holds with probability $0.99$ \cite[Theorem 7]{woodruff2014sketching}.
\end{itemize}

%---------------------------------Figure---------------------------------%
\begin{figure}[!ht]
\begin{center}
\centering
\subfigure[\textsf{Hash each column with a value uniformly sampled from $[s]=\{1,2,3\}$.}]{\includegraphics[width=0.85\textwidth]{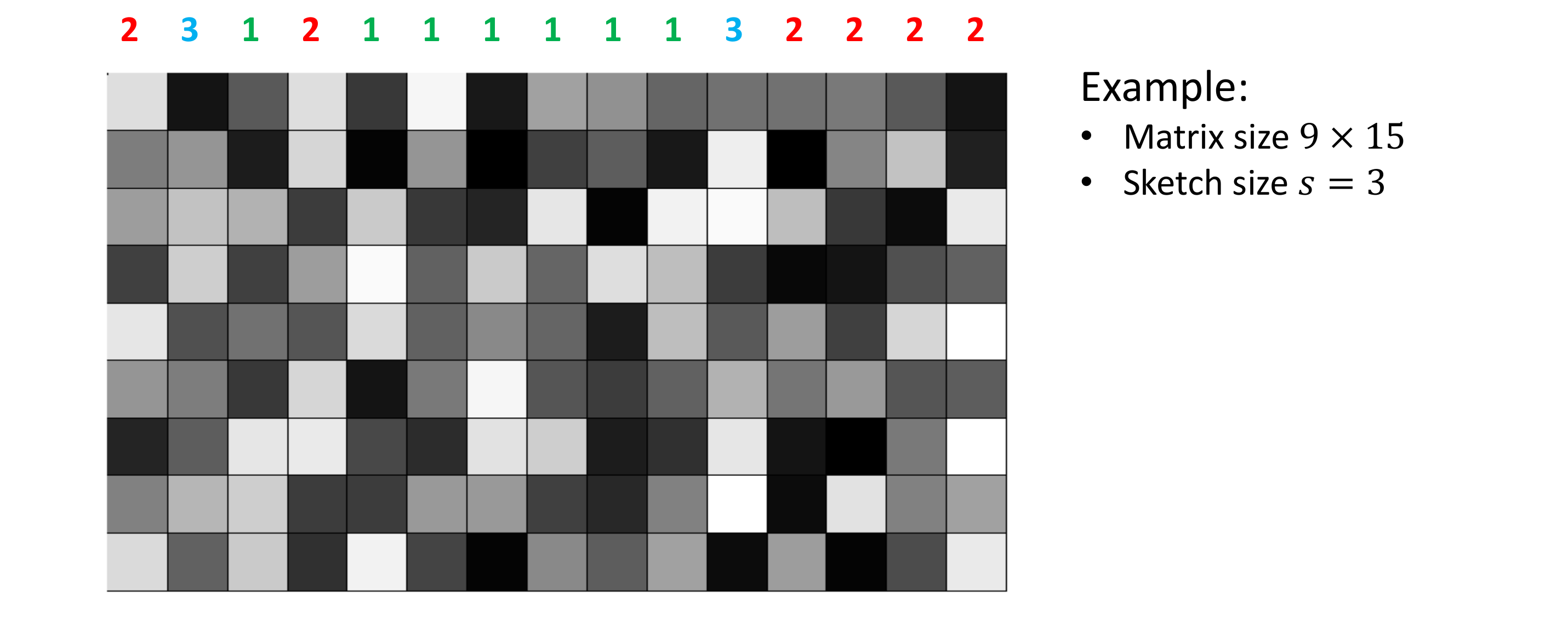}}
\subfigure[\textsf{Flip the sign of each column with probability $50\%$, and then sum up columns with the same hash value.}]{\includegraphics[width=0.88\textwidth]{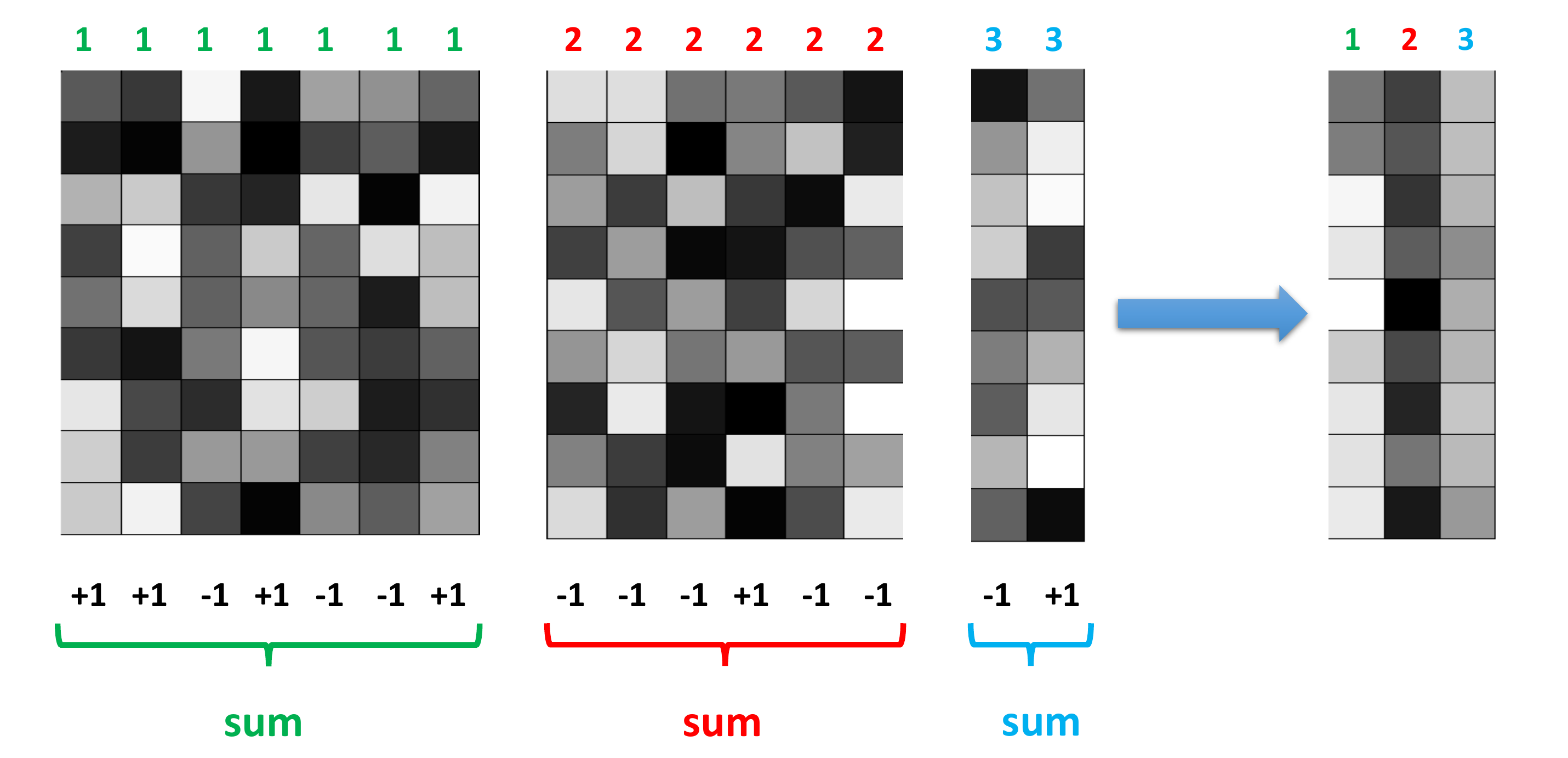}}
\end{center}
   \caption{Count sketch in the map-reduce fashion.}
\label{fig:count_sketch_mr}
\end{figure}
%---------------------------------Figure---------------------------------%

\subsection{Count Sketch} \label{sec:countsketch}

Count sketch stems from the data stream literature \citep{charikar2004finding,thorup2012tabulation}.
It was applied to speedup matrix computation by \cite{clarkson2013low,pham2013fast}.
We describe in the following the count sketch for matrix data.

\begin{figure}[!ht]
\begin{center}
\centering
\includegraphics[width=0.85\textwidth]{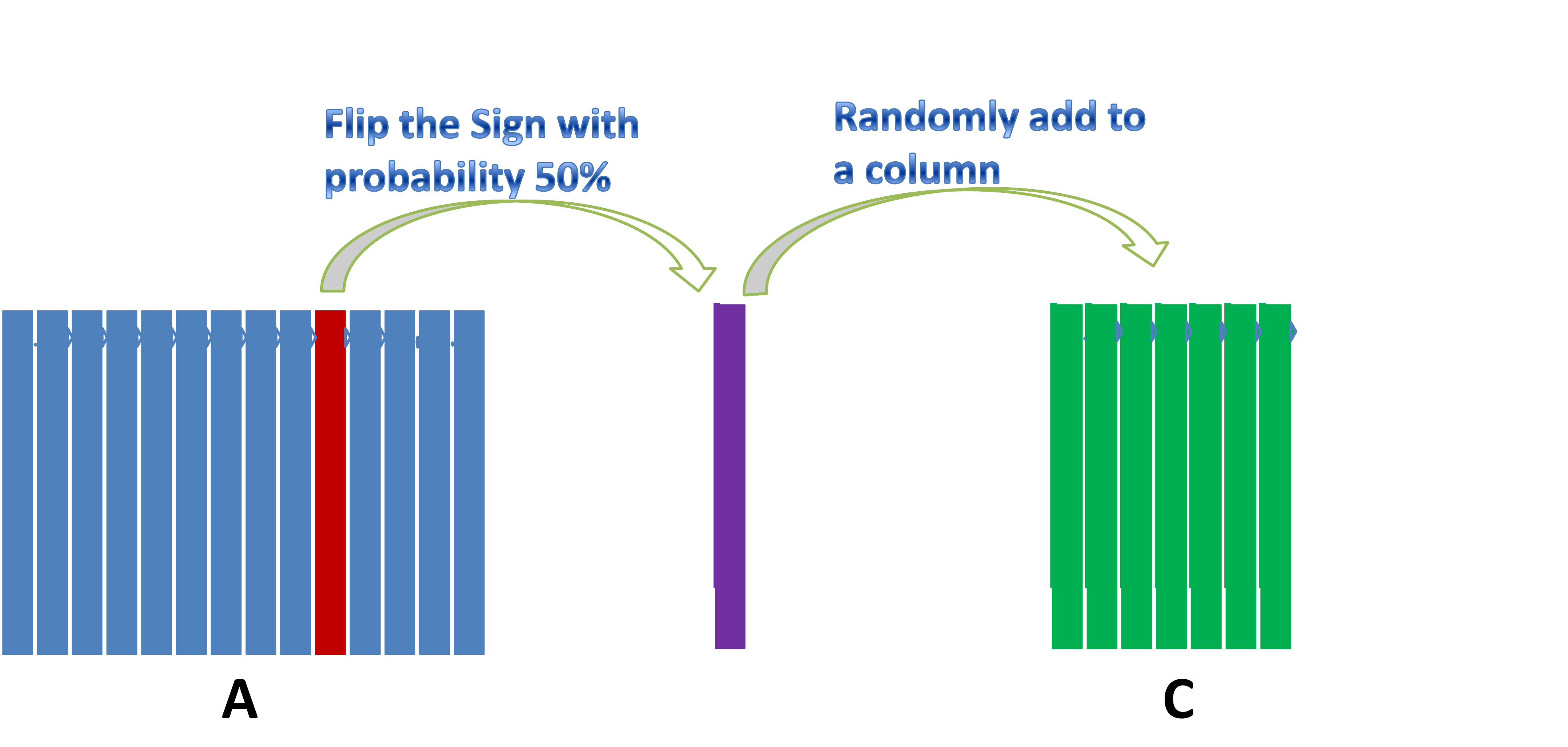}
\end{center}
   \caption{Count sketch in the streaming fashion.}
\label{fig:count_sketch_streaming}
\end{figure}

There are different ways to implementing count sketch.
This paper describe two quite different ways and refer to them as ``map-reduce fashion'' and ``streaming fashion''.
Of course, the two are equivalent.
\begin{itemize}
\item
    The map-reduce fashion has three steps.
    First, hash each column with a discrete value uniformly sampled from $[s]$.
    Second, flip the sign of each column with probability $50\%$.
    Third, sum up columns with the same hash value.
    This procedure is illustrated in Figure~\ref{fig:count_sketch_mr}.
    As its name suggests, this approach naturally fits the map-reduce systems.
\item
    The streaming fashion has two steps.
    First, initialize $\C$ to be the $m\times s$ all-zero matrix.
    Second, for each column of $\A$, flip its sign with probability $50\%$, and add it to a uniformly selected column of $\C$.
    It is described in Algorithm~\ref{alg:count_sketch} an illustrated in Figure~\ref{fig:count_sketch_streaming}.
    It can be implemented in 9 lines of MATLAB code as below.
    The streaming fashion implementation keeps the sketch $\C$ in memory and scans the data $\A$ in only one pass.
    If $\A$ does not fit in memory, this approach is better than the map-reduce fashion for it scans the columns sequentially.
    If $\A$ is sparse matrix, randomly accessing the entries may not be efficient,
    and thus it is better to accessing the column sequentially.
\end{itemize}

\vspace{3mm}
\begin{lstlisting}
function [C] = CountSketch(A, s) % the streaming fashion
[m, n] = size(A);
sgn = randi(2, [1, n]) * 2 - 3; % one half are +1 and the rest are -1
A = bsxfun(@times, A, sgn); % flip the signs of each column w.p. 50%
ll = randsample(s, n, true); % sample n items from [s] with replacement
C = zeros(m, s); % initialize C
for j = 1: n
    C(:, ll(j)) = C(:, ll(j)) + A(:, j);
end
\end{lstlisting}

\begin{algorithm}[!t]
   \caption{Count Sketch in the Streaming Fashion.}
   \label{alg:count_sketch}
\algsetup{indent=2em}
%\begin{small}
\begin{algorithmic}[1]
   \STATE {\bf input:} $\A \in \RB^{m\times n}$.
   \STATE Initialize $\C$ to be an $m\times s$ all-zero matrix;
   \FOR{$i = 1$ to $n$}
   \STATE sample $l$ from the set $[s]$ uniformly at random;
   \STATE sample $g$ from the set $\{ +1 , -1\}$ uniformly at random;
   \STATE update the $l$-th column of $\C$ by $\c_{:l} \longleftarrow \c_{:l} + g \a_{:i}$;
   \ENDFOR
   \RETURN $\C \in \RB^{m\times s}$.
\end{algorithmic}
%\end{small}
\end{algorithm}

%---------------------------------Figure---------------------------------%
%---------------------------------Figure---------------------------------%

The readers may have noticed that count sketch does not explicitly form the sketching matrix $\S$.
In fact, $\S$ is such a matrix that each row has only one nonzero entry.
In the example of Figure~\ref{fig:count_sketch_mr}, the matrix $\S^T$ can be explicitly expressed as

\begin{footnotesize}
\[
\S^T \; = \;
\left[
  \begin{array}{ccccccccccccccc}
    0 & 0 & 1 & 0 & 1 &-1 & 1 &-1 &-1 & 1 & 0 & 0 & 0 & 0 & 0 \\
    -1& 0 & 0 & -1& 0 & 0 & 0 & 0 & 0 & 0 & 0 & -1& 1 & -1&-1 \\
    0 &-1 & 0 & 0 & 0 & 0 & 0 & 0 & 0 & 0 & 1 & 0 & 0 & 0 & 0 \\
  \end{array}
\right].
\]%
\end{footnotesize}%

\vspace{3mm}
Count sketch has the following properties:

\begin{itemize}
\item
    Time cost: $\OM (\nnz (\A))$
\item
    Memory cost: $\OM ( m s)$.
    When $\A$ does not fit in memory, the algorithm keeps only $\C$ in memory and goes one pass through the columns of $\A$.
\item
    Theoretical guarantees
    \begin{enumerate}
    \item
        When $s = \OM ( {m^2}/{\epsilon^2 } )$,
        the subspace embedding property holds with $\gamma = 1+\epsilon$ with high probability.
    \item
        When $s = \OM ( {k}/{\epsilon } + k^2 )$, the low-rank approximation property holds with $\eta = 1+\epsilon$ relative error with high probability.
    \end{enumerate}
\item
    Advantage: the count sketch is very efficient, especially when $\A$ is sparse.
\item
    Disadvantage: compared with Gaussian projection, the count sketch requires larger $s$ to attain the same accuracy.
        One simple improvement is to combine the count sketch with Gaussian projection or SRHT.
\end{itemize}

\subsection{GaussianProjection + CountSketch} \label{sec:sketching:gaussian_countsketch}

Let $\S_{sc}$ be $n\times s_{cs}$ count sketch matrix, $\S_{gp}$ be $s_{cs} \times s$ Gaussian projection matrix,
and $\S = \S_{cs} \S_{gp} \in \RB^{n\times s}$.
Then $\S$ satisfies the following properties.
\begin{itemize}
\item
    Time complexity: the matrix product $\A \S$ can be computed in
    \[
    \OM \big( \underbrace{\nnz (\A)}_{\textrm{count sketch}} +\underbrace{ m s_{cs} s}_{\textrm{Gaussian projection}} \big)
    %\; = \;\tilde\OM \Big( \nnz (\A) + m^4 \epsilon^{-4}  \Big)
    \]
    time.
\item
    Theoretical properties:
    \begin{enumerate}
    \item
        When $s_{cs} = \OM (m^2 / \epsilon^{2})$ and $s = \OM( m /\epsilon^{2} )$,
        the GaussianProjection+CountSketch matrix $\S$ satisfy the subspace embedding property with $\gamma = 1+\epsilon$ holds with high probability.
    \item
        When $s_{cs} = \OM (k^2 + k/\epsilon)$ and $s = \OM( k / \epsilon )$,
        the GaussianProjection+CountSketch matrix $\S$ satisfies the low-rank approximation property with $\eta = 1+\epsilon$
        \cite[Lemma 12]{boutsidis2015communication}.
    \end{enumerate}
\item
    Advantages:
    \begin{enumerate}
    \item
        the size of GaussianProjection+CountSketch is as small as Gaussian projection.
    \item
        the time complexity is much lower than Gaussian projection when $n \gg m$.
    \end{enumerate}
\end{itemize}

\section{Column Selection} \label{sec:columnselection}

This section presents three column selection techniques: uniform sampling, leverage score sampling, and local landmark selection.
Different from random projection, column selection do not have to visit every entry of $\A$,
and column selection preserves the sparsity/non-negativity properties of $\A$.

\subsection{Uniform Sampling} \label{sec:columnselection:uniform}

Uniform sampling is the most efficient way to form a sketch.
The most important advantage is that uniform sampling forms a sketch without seeing the whole data matrix.
When applied to kernel methods, uniform sampling avoids computing every entry of the kernel matrix.

The performance of uniform sampling is data-dependent.
When the leverage scores (defined in Section~\ref{sec:columnselection:leverage}) are uniform,
or equivalently, the matrix coherence (namely the greatest leverage score) is small,
uniform sampling has good performance.
The analysis of uniform sampling can be found in \cite{gittens2011spectral,gittens2013revisiting}.

\subsection{Leverage Score Sampling} \label{sec:columnselection:leverage}

Before studying leverage score sampling, let's first define leverage scores.
Let $\A$ be an $m\times n$ matrix, with $\rho = \rk(\A) < n$,
and $\V \in \RB^{n\times \rho}$ be the right singular vectors.
The (column) leverage scores of $\A$ are defined by
\[
l_i \; := \; \|\v_{i:}\|_2^2, \qquad \textrm{ for } i = 1 , \cdots ,n.
\]
Leverage score sampling is to select each columns of $\A$ with probability proportional to its leverage scores.
(Sometimes each selected column should be scaled by $\sqrt{\frac{\rho}{s l_i}}$.)
It can be roughly implemented in 8 lines MATLAB code.

\vspace{3mm}

\begin{lstlisting}
function [C, idx] = LeverageScoreSampling(A, s)
n = size(A, 2);
[~, ~, V] = svd(A, 'econ');
leveragescores = sum(V.^2, 2);
prob = leveragescores / sum(leveragescores);
idx = randsample(n, s, true, prob);
idx = unique(idx); % eliminate duplicates
C = A(:, idx);
\end{lstlisting}
\vspace{3mm}
There are a few things to remark:
\begin{itemize}
\item
    To sample columns according to the leverage scores of $\A_k$ where $k \ll m, n$,
    Line 3 can be replaced by
\begin{lstlisting}[firstnumber=3]
[~, ~, V] = svds(A, k);
\end{lstlisting}
\item
    Theoretical properties
    \begin{enumerate}
    \item
        When $s = \OM ( m / \epsilon + m \log m  )$,
        the leverage score sampling satisfies the subspace embedding property with $\gamma = 1+\epsilon$ holds with high probability.
    \item
        When $s = \OM ( k / \epsilon + k \log k)$,
        the leverage score sampling (according to the leverage scores of $\A_k$) satisfies the low-rank approximation property with $\eta = 1+\epsilon$.
    \end{enumerate}
\item
    Computing the leverage scores is as expensive as computing SVD,
    so leverage score sampling is not a practical way to sketch the matrix $\A$ itself.
\item
    When the leverage scores are near uniform,
    there is little difference between uniform sampling and leverage score sampling.
\end{itemize}

\subsection{Local Landmark Selection} \label{sec:columnselection:landmark}

Local landmark selection is a very effective heuristic for finding representative columns.
\citet{zhang2010clustered} proposed to set $k=s$ and run $k$-means or $k$-centroids clustering algorithm to cluster the columns of $\A$ to $s$ class,
and use the $s$ centroids as the sketch of $\A$.
This heuristic works very well in practice, though it has little theoretical guarantee.

There are several tricks to make the local landmark selection more efficient.
\begin{itemize}
\item
    One can simply solve $k$-centroids clustering approximately rather than accurately.
    For example, it is unnecessary to wait for $k$-centroids clustering to converge;
    running $k$-centroids for a few iterations suffices.
\item
    When $n$ is large, one can uniformly sample a subset of the data, e.g.\ $\max\{0.2n , 20s\}$ data points,
    and perform local landmark selection on this smaller dataset.
\item
    In supervised learning problems, each datum $\a_i$ is associated with a label $y_i$.
    We can partition the data to $g$ groups according to the labels and run $k$-centroids clustering independently on the data in each group.
    In this way, $s = gk$ data points are selected as a sketch of $\A$.
\end{itemize}

%%%%%%%%%%%%%%%%%%%%%%%%%%%%%%%%%%%%%%%%%%%%%%%%%%%%%%%%%%%%%%%%%%%%%%%%%%%%%%
%%%%%%%%%%%%%%%%%%%%%%%%%%%%%%%%%%%%%%%%%%%%%%%%%%%%%%%%%%%%%%%%%%%%%%%%%%%%%%
%%%%%%%%%%%%%%%%%%%%%%%%%%%%%%%%%%%%%%%%%%%%%%%%%%%%%%%%%%%%%%%%%%%%%%%%%%%%%%
%%%%%%%%%%%%%%%%%%%%%%%%%%%%%%%%%%%%%%%%%%%%%%%%%%%%%%%%%%%%%%%%%%%%%%%%%%%%%%

\chapter{Regression} \label{sec:lsr}

Let $\A$ be an $n\times d$ ($n\geq d$) matrix whose rows correspond to data and columns correspond to features,
and let $\bb \in \RB^{n}$ contain the response/label of each datum.
The least squares regression (LSR)
\begin{equation} \label{eq:lsr_def}
\min_\x \; \|\A \x - \bb \|_2^2
\end{equation}
is a ubiquitous problem in statistics, computer science, economics, etc.
When $n \gg d$, LSR can be efficiently solved using randomized algorithms.

\section{Standard Solutions} \label{sec:lsr_standard}

The least squares regression (LSR) problem \eqref{eq:lsr_def} has closed form solution
\begin{equation}
\x^\star \; = \; \A^\dag \bb  . \nonumber
\end{equation}
The Moore-Penrose inverse can be computed by SVD which costs $\OM (n d^2)$ time.

LSR can also be solved by numerical algorithms such as the conjugate gradient (CG) algorithm,
and machine-precision can be attained in a reasonable number of iterations.
Let $\kappa (\A) := \frac{\sigma_1 (\A)}{\sigma_d (\A)}$ be the condition number of $\A$.
The convergence of CG depends on $\kappa (\A)$:
\[
\frac{\|\A (\x^{(t)} - \x^\star)\|_2^2}{\|\A (\x^{(0)} - \x^\star)\|_2^2}
\; \leq \;
2 \bigg( \frac{ \kappa (\A) -1  }{ \kappa (\A) + 1 } \bigg)^t ,
\]
where $\x^{(t)}$ is the model in the $t$-th iteration of CG.
The per-iteration time cost of CG is $\OM(\nnz (\A))$.
To attain $\|\A (\x^{(t)} - \x^\star)\|_2^2 \leq \epsilon$,
the number of iteration is roughly
\[
\Big( \log \frac{1}{\epsilon} + \log (\textrm{InitialError}) \Big)  \frac{ \kappa (\A) - 1}{2} .
\]
Since the time cost of CG heavily depends on the unknown condition number $\kappa (\A)$,
CG can be very slow if $\A$ is ill-conditioned.

\section{Inexact Solution} \label{sec:lsr:inexact}

Any sketching matrix $\S \in \RB^{n\times s}$ can be used to solve LSR approximately as long as it satisfies the subspace embedding property.
We consider the following LSR problem:
\begin{equation} \label{eq:lsr_inexact}
\tilde\x \; = \; \min_\x \; \| \underbrace{(\S^T \A)}_{s\times d} \x - \S^T \bb \|_2^2 ,
\end{equation}
which can be solved in $\OM (s d^2)$ time.

If $\S$ is a Gaussian projection matrix, SRHT matrix, count sketch, or leverage score sampling matrix,
and $s = \poly (d / \epsilon)$ for any error parameter $\epsilon \in (0, 1]$,
then
\[
\| \A \tilde\x - \bb\|_2^2 \; \leq \; (1+\epsilon)^2 \min_\x \| \A \x - \bb \|_2^2
\]
is guaranteed.

\subsection{Implementation}

If $\S$ is count sketch matrix, the inexact LSR algorithm can be implemented in 5 lines of MATLAB code.
Here CountSketch is a MATLAB function described in Section~\ref{sec:countsketch}.
The total time cost is $\OM (\nnz(\A) + \poly (d /\epsilon))$ and memory cost is $\OM (\poly (d/\epsilon))$,
which are lower than the cost of exact LSR when $d \ll n$.

\vspace{3mm}
\begin{lstlisting}
function [xtilde] = InexactLSR(A, b, s)
d = size(A, 2);
sketch = (CountSketch([A, b]', s))';
Asketch = sketch(:, 1:d); % Asketch = S' * A
bsketch = sketch(:, end); % bsketch = S' * b
xtilde = Asketch \ bsketch;
\end{lstlisting}
\vspace{3mm}
There are a few things to remark:
\begin{itemize}
\item
    The inexact LSR is useful only when $n = \Omega (d/\epsilon + d^2)$.
\item
    The size of sketch $s$ is a polynomial function of $\epsilon^{-1}$ rather than logarithm of $\epsilon^{-1}$,
    thus the algorithm cannot attain high precision.
\end{itemize}

\subsection{Theoretical Explanation}

By the subspace embedding property,
it can be easily shown that $\tilde\x$ is a good solution.
Let $\D = [\A, \bb] \in \RB^{n\times (d+1)}$ and $\z = [\x; -1] \in \RB^{n+1}$.
Then
\[
\A \x - \bb = \D \z
\qquad\textrm{ and }\qquad
\S^T \A \x - \S^T \bb = \S^T \D \z ,
\]
and the subspace embedding property indicates
$\frac{1}{\eta} \| \D \z \|_2^2 \leq \| \S^T \D \z \|_2^2 \leq \eta  \| \D \z \|_2^2 $ for all $\z$.
Thus
\[
\frac{1}{\eta} \| \A \tilde\x - \bb\|_2^2
\; \leq \; \| \S^T (\A \tilde\x - \bb) \|_2^2
\qquad\textrm{ and }\qquad
\| \S^T (\A \x^\star - \bb) \|_2^2
\; \leq \; \eta \| \A \x^\star - \bb \|_2^2
\]
The optimality of $\tilde\x$ indicates $\| \S^T (\A \tilde\x - \bb) \|_2^2 \leq \| \S^T (\A \x^\star - \bb) \|_2^2$,
and thus
\begin{align*}
&\frac{1}{\eta} \| \A \tilde\x - \bb\|_2^2
\; \leq \; \| \S^T (\A \tilde\x - \bb) \|_2^2
\; \leq \; \| \S^T (\A \x^\star - \bb) \|_2^2
\; \leq \; \eta \| \A \x^\star - \bb \|_2^2 . \\
&\Rightarrow
\| \A \tilde\x - \bb\|_2^2 \; \leq \; \eta^2 \| \A \x^\star - \bb \|_2^2.
\end{align*}
Therefore, as long as $\S$ satisfies the subspace embedding property,
the approximate solution to LSR is nearly as good as the optimal solution (in terms of objective function value).

\section{Machine-Precision Solution} \label{sec:lsr:machine_precision}

Randomized algorithms can also be applied to find machine-precision solution to LSR,
and the time complexity is lower than the standard solutions.
The state-of-the-art algorithm \cite{meng2014lsrn} is based on very similar idea described in this section.

\subsection{Basic Idea: Preconditioning}

We have discussed previously that the time cost of the conjugate gradient (CG) algorithm is roughly
\[
 \frac{ \red{\kappa (\A)} - 1}{2} \Big( \log \frac{1}{\epsilon} + \log (\textrm{InitialError}) \Big) \nnz (\A) ,
\]
which dependents on the condition number of $\A$.
To make CG efficient, one can find a $d\times d$ preconditioning matrix $\T$ such that $\kappa (\A \T)$ is small,
solve
\begin{equation} \label{eq:lsr_machine_precision}
\z^\star \; = \; \argmin_\z \| (\A \T) \z - \bb\|_2^2
\end{equation}
by CG, and let $\x^\star = \T \z^\star$.
In this way, the time cost of CG is roughly
\[
 \frac{ \red{\kappa (\A \T)} - 1}{2} \Big( \log \frac{1}{\epsilon} + \log (\textrm{InitialError}) \Big) \nnz (\A) .
\]
If $\kappa (\A \T)$ is a small constant, e.g.\ $\kappa (\A \T) = 2$,
then \eqref{eq:lsr_machine_precision} can be very efficiently solved by CG.

Now let's consider how to find the preconditioning matrix $\T$.
Let $\A = \Q_\A \R_\A$ be the QR decomposition.
Obviously $\T = \R_\A^{-1}$ is a perfect preconditioning matrix because $\kappa (\A \R_\A^{-1}) = \kappa (\Q_\A) = 1$.
Unfortunately, the preconditioning matrix $\T = \R_\A^{-1}$ is not a practical choice because computing the QR decomposition is as expensive as solving LSR.

Woodruff \cite{woodruff2014sketching} proposed to use sketching to find $\R_\A$ approximately in $\OM (\nnz (\A) + \poly (d))$ time.
Let $\S \in \RB^{n\times s}$ be a sketching matrix and form $\Y = \S^T \A$.
Let $\Y = \Q_\Y \R_\Y$ be the QR decomposition of $\Y$.
Theory shows that the sketch size $s = \OM (d^2)$ suffices for $\kappa (\A \R_\Y^{-1}) \leq  2$ holding with high probability.
Thus $\R_\Y^{-1} \in \RB^{d\times d}$ is a good preconditioning matrix.

\begin{algorithm}[!t]
   \caption{Machine-Precision Solution to LSR.}
   \label{alg:lsr}
\algsetup{indent=2em}
%\begin{small}
\begin{algorithmic}[1]
   \STATE {\bf input:} $\A \in \RB^{n\times d}$, $\bb \in \RB^n$, and step size $\theta$.
   \STATE Draw a sketching matrix $\S \in \RB^{n\times s}$ where $s = \OM (d^2)$;
    \STATE Form the sketch $\Y = \S^T \A \in \RB^{s\times d}$;
    \STATE Compute the QR decomposition $\Y = \Q_\Y \R_\Y$;
    \STATE Compute the preconditioning matrix $\T = \R_\Y^{-1}$;
    \STATE Compute the initial solution $\z^{(0)} = (\S^T \A \T)^\dag (\S^T \bb) = \Q_\Y^T (\S^T \bb)$;
    \FOR{$t = 1, \cdots , \OM (\log \epsilon^{-1})$}
        \STATE $\rr^{(t)} = \bb - \A \T \z^{(t-1)}$ ;  \quad\qquad\; // the residual \label{alg:lsr:gradient1}
        \STATE $\z^{(t)} = \z^{(t-1)} + \theta \T^T \A^T \rr^{(t)}$;   \qquad // gradient descent  \label{alg:lsr:gradient2}
    \ENDFOR
    \RETURN $\x^\star = \T \z^{(t)} \in \RB^{d}$.
\end{algorithmic}
%\end{small}
\end{algorithm}

\subsection{Algorithm Description}

The algorithm is described in Algorithm~\ref{alg:lsr}.
We first form a sketch $\Y = \S^T \A \in \RB^{s\times d}$ and compute its QR decomposition $\Y = \Q_\Y \R_\Y$.
We can use this QR decomposition to find the initial solution $\z^{(0)}$ and the preconditioning matrix $\T = \R_\Y^{-1}$.
If we set $s = \OM (d^2)$, the initial solution is only constant times worse than the optimal in terms of objective function value.
Theory also ensures that the condition number $\kappa (\A \T) \leq 2$.
With the good initialization and good condition number, the vanilla gradient descent\footnote{Since $\A\T$ is well conditioned, the vanilla gradient descent and CG has little difference.}
or CG takes only $\OM (\log \epsilon^{-1})$ steps to attain $1+\epsilon$ solution.
Notice that Lines~\ref{alg:lsr:gradient1} and \ref{alg:lsr:gradient2} in the algorithm should be cautiously implemented.
Do not compute the matrix product $\A \T$ because it would take $\OM (\nnz(\A) d)$ time!

\section{Extension: CX-Type Regression} \label{sec:lsr:extension1}

Given any matrix $\A \in \RB^{m\times n}$,
CX decomposition considers decomposing $\A$ into $\A \approx \C \X^\star$,
where $\C \in \RB^{m\times c}$ is a sketch of $\A$ and $\X^\star \in \RB^{c\times n}$ is computed by
\[
\X^\star
\; = \; \argmin_\X \, \big\|\A - \C \X \big\|_F^2
\; = \; \C^\dag \A .
\]
It takes $\OM (m n c)$ time to compute $\X^\star$.
If $c \ll m$, this problem can be solved more efficiently by sketching.
Specifically, we can draw a sketching matrix $\S \in \RB^{m\times s}$ and compute the approximate solution
\[
\tilde\X \; = \; \argmin_\X \|\underbrace{\S^T\C}_{s\times c} \underbrace{\X}_{c\times n} - \underbrace{\S^T\A}_{s\times n}\|_F^2
\; = \; (\S^T \C)^\dag (\S^T \A)
\]
If $\S$ is a count sketch matrix, we set $s = \OM (c/\epsilon + c^2)$;
if $\S$ samples columns according to the row leverage scores of $\C$, we set $s = \OM (c / \epsilon + c \log c)$.
It holds with high probability that
\[
\big\|\A - \C \tilde\X \big\|_F^2
\; \leq \; (1+\epsilon) \, \min_\X \big\|\A - \C \X \big\|_F^2.
\]

\section{Extension: CUR-Type Regression} \label{sec:lsr:extension2}

A more complicated problem has also been considered in the literature \cite{stewart1999four,wang2013improving,si2014memory}:
\begin{equation} \label{eq:general_lsr}
\X^\star \; = \; \argmin_\X \| \underbrace{\C}_{n\times c} \underbrace{\X}_{c\times r} \underbrace{\R}_{r\times n} - \underbrace{\A}_{m\times n} \|_F^2
\end{equation}
where $c, r \ll m, n$.
The solution is:
\[
\X^\star \; = \; \C^\dag \A \R^\dag  ,
\]
which cost $\OM (m n \cdot \min\{c, r\})$ time.
\citet{wang2015towards} proposed an algorithm to solve \eqref{eq:general_lsr} approximately by
\[
\tilde\X \; = \; \argmin_\X \| \S_C^T (\C \X \R - \A )\S_R \|_F^2
\]
where $\S_C \in \RB^{m\times s_c}$ and $\S_R \in  \RB^{n\times s_r}$ are leverage score sampling matrices.
When $s_c = c \sqrt{q/\epsilon}$ and $s_r = r \sqrt{q/\epsilon}$ (where $q = \min\{m, n\}$),
it holds with high probability that
\[
\| \C \tilde\X \R - \A  \|_F^2
\; \leq \; (1+\epsilon) \, \min_\X \| \C \X \R - \A  \|_F^2.
\]
The total time cost is
\[
\OM (s_c s_r \cdot \min \{c, r\}) \; = \; \OM (c r \epsilon^{-1} \cdot \min \{m, n\} \cdot \min \{c, r\})
\]
time,
which is useful when $\max\{m , n\} \gg c, r$.
The algorithm can be implemented in 4 lines of MATLAB code:
\vspace{3mm}
\begin{lstlisting}
function [Xtilde] = InexactCurTypeRegression(C, R, A, sc, sr)
[~, idxC] = LeverageScoreSampling(C', sc);
[~, idxR] = LeverageScoreSampling(R, sr);
Xtilde = pinv(C(idxC, :)) * A(idxC, idxR) * pinv(R(:, idxR));
\end{lstlisting}
\vspace{3mm}
Here the function ``$\mathrm{LeverageScoreSampling}$'' is described in Section~\ref{sec:columnselection:leverage}.
Empirically, setting $s_1 = s_2 = \OM (d_1 + d_2)$ suffices for high precision.
The experiments in \cite{wang2015towards} indicates that uniform sampling performs equally well as leverage score sampling.

%%%%%%%%%%%%%%%%%%%%%%%%%%%%%%%%%%%%%%%%%%%%%%%%%%%%%%%%%%%%%%%%%%%%%%%%%%%%%%
%%%%%%%%%%%%%%%%%%%%%%%%%%%%%%%%%%%%%%%%%%%%%%%%%%%%%%%%%%%%%%%%%%%%%%%%%%%%%%
%%%%%%%%%%%%%%%%%%%%%%%%%%%%%%%%%%%%%%%%%%%%%%%%%%%%%%%%%%%%%%%%%%%%%%%%%%%%%%
%%%%%%%%%%%%%%%%%%%%%%%%%%%%%%%%%%%%%%%%%%%%%%%%%%%%%%%%%%%%%%%%%%%%%%%%%%%%%%

\chapter{Rank $k$ Singular Value Decomposition}  \label{sec:ksvd}

This chapter considers the $k$-SVD of a large scale matrix $\A \in \RB^{m\times n}$,
which may not fit in memory.

\section{Standard Solutions} \label{sec:ksvd_standard}

The standard solutions to $k$-SVD include the power iteration algorithm and the Krylov subspace methods.
Their time complexities are considered to be $\tilde\OM (m n k)$,
where the $\tilde\OM$ notation hides parameters such as the spectral gap and logarithm of error tolerance.
Here we introduce a simplified version of the block Lanczos method \cite{musco2015stronger}\footnote{We introduce this
algorithm because it is easy to understand. However, as $q$ grows, columns of the Krylov matrix gets increasingly linearly dependent,
which sometimes leads to instability.
Thus there are many numerical treatments to strengthen stability (see the numerically stable algorithms in \cite{saad2011numerical}).}
which costs time $\OM (m n k q)$,
where $q = \log \frac{n}{\epsilon}$ is the number of iterations, and the inherent constant depends weakly on the spectral gap.
The block Lanczos algorithm is described in Algorithm~\ref{alg:block_lanczos} can be implemented in 18 lines of MATLAB code.

\vspace{3mm}

\begin{lstlisting}
function [U, S, V] = BlockLanczos(A, k, q)
s = 2 * k; % can be tuned
[m, n] = size(A);
C = A * randn(n, s);
Krylov = zeros(m, s * q);
Krylov(:, 1:s) = C;
for i = 2: q
    C = A' * C;
    C = A * C;
    [C, ~] = qr(C, 0); % optional
    Krylov(:, (i-1)*s+1: i*s) = C;
end
[Q, ~] = qr(Krylov, 0);
[Ubar, S, V] = svd(Q' * A, 'econ');
Ubar = Ubar(:, 1:k);
S = S(1:k, 1:k);
V = V(:, 1:k);
U = Q * Ubar;
\end{lstlisting}
\vspace{3mm}
Although the block Lanczos algorithm can attain machine precision,
it inevitably goes many passes through $\A$, and it is thus slow when $\A$ does not fit in memory.

Facing large-scale data, we must trade off between precision and computational costs.
We are particularly interested in approximate algorithm that satisfies:
\begin{enumerate}
\item
    The algorithm goes constant passes through $\A$.
    Then $\A$ can be stored in large volume disks, and there are only constant swaps between disk and memory.
\item
    The algorithm only keeps a small-scale sketch of $\A$ in memory.
\item
    The time cost is $\OM (m n k)$ or lower.
\end{enumerate}

\begin{algorithm}[tb]
    \caption{$k$-SVD by the Block Lanczos Algorithm.}
    \label{alg:block_lanczos}
\begin{small}
\begin{algorithmic}[1]
    \STATE {\bf Input:} an $m\times n$ matrix $\A$ and the target rank $k$.
    \STATE Set $s = k + \OM (1)$ be the over-sampling parameter;
    \STATE Set $q = \OM (\log \frac{n}{\epsilon})$ be the number of iteration;
    \STATE Draw a $n\times s$ sketching matrix $\S$;
    \STATE $\C = \A \S$;
    \STATE Set $\K = \big[\C ,\; (\A \A^T)\C ,\; (\A \A^T)^2 \C ,\; \cdots , \; (\A \A^T)^{q-1} \C \big]$;
    \STATE QR decomposition: $[\underbrace{\Q_\C}_{m\times sq} , \R_\C] = qr(\underbrace{\K}_{m\times sq})$;
    \STATE SVD: $[\underbrace{\bar\U}_{sq\times sq} , \underbrace{\Si}_{sq\times sq} , \underbrace{\V}_{n\times sq}]
            = svd(\underbrace{\Q_\C^T \A}_{s\times n})$;
    \STATE Retain the top $k$ components of $\bar\U$, $\Si$, and $\V$ to form $sq\times k$, $k\times k$, $n\times k$ matrices;
    \STATE $\U = \Q \bar\U \in \RB^{m\times k}$;
    \RETURN $\U \Si \V^T \approx \A_k$.
\end{algorithmic}
\end{small}
\end{algorithm}

\section{Prototype Randomized $k$-SVD Algorithm} \label{sec:ksvd_prototype}

This section describes a randomized algorithm that computes the $k$-SVD of $\A$ up to $1+\epsilon$ Frobenius norm relative error.
The algorithm is proposed by \cite{halko2011ramdom},
and it is described in Algorithm~\ref{alg:ksvd_prototype}.

\begin{algorithm}[tb]
    \caption{Prototype Randomized $k$-SVD Algorithm.}
    \label{alg:ksvd_prototype}
\begin{small}
\begin{algorithmic}[1]
    \STATE {\bf Input:} an $m\times n$ matrix $\A$ and the target rank $k$.
    \STATE Draw a $n\times s$ sketching matrix $\S$ where $s = \OM(\frac{k}{\epsilon})$;
    \STATE $\C = \A \S$;
    \STATE QR decomposition: $[\underbrace{\Q_\C}_{m\times s} , \R_\C] = qr(\underbrace{\C}_{m\times s})$;
    \STATE $k$-SVD: $[\underbrace{\bar\U}_{s\times k} , \underbrace{\tilde\Si}_{k\times k} , \underbrace{\tilde\V}_{n\times k}]
            = svds(\underbrace{\Q_\C^T \A}_{s\times n}, k)$;
    \STATE $\tilde\U = \Q_\C \bar\U \in \RB^{m\times k}$;
    \RETURN $\tilde\U \tilde\Si \tilde\V^T \approx \A_k$.
\end{algorithmic}
\end{small}
\end{algorithm}

\subsection{Theoretical Explanation}

If $\C = \A \S \in \RB^{m\times s}$ is a good sketch of $\A$, the column space of $\C$ should roughly
contain the columns of $\A_k$---this is the low-rank approximation property.
If $\S\in \RB^{n\times s}$ is Gaussian projection matrix or count sketch
and $s = \OM (k/\epsilon)$,
then the low-rank approximation property
\begin{equation}\label{eq:ksvd_prototype_C}
\min_{\rk (\Z) \leq k} \| \C \Z  - \A \|_F^2
\; \leq \; (1+\epsilon) \| \A - \A_k \|_F^2
\end{equation}
holds in expectation.

\subsection{Algorithm Derivation}

Let $\Q_\C$ be any orthonormal bases of $\C$.
Since the column space of $\C$ is the same to the column space of $\Q_\C$,
the minimization problem in \eqref{eq:ksvd_prototype_C} can be equivalently converted to
\begin{equation}\label{eq:ksvd_prototype}
\X^\star \; = \; \argmin_{\rk(\X) \leq k} \| \underbrace{\Q_\C}_{m\times s} \underbrace{\X}_{s\times n} - \underbrace{\A }_{m\times n} \|_F^2
\; = \; (\Q_\C^T \A)_k .
\end{equation}
Here the second equality is a well known fact.
The matrix $\A_k$ is well approximated by $\tilde\A_k : = \Q_\C \X^\star $,
so we need only to find the $k$-SVD of $\tilde\A_k$:
\begin{eqnarray*}
\tilde\A_k \; := \;
\underbrace{\Q_\C}_{m\times s} \underbrace{\X^\star }_{s\times n}
\; = \; \Q_\C \underbrace{(\Q_\C^T \A)_k}_{:= \bar\U {\tilde{\bf \Sigma}} \tilde\V^T}
\; = \; \underbrace{\Q_\C \bar\U}_{:= \tilde\U} \tilde\Si \tilde\V^T
\; = \; \underbrace{\tilde\U}_{m\times k} \underbrace{\tilde\Si}_{k\times k} \underbrace{\tilde\V^T}_{k\times n} .
\end{eqnarray*}
It is easy to check that $\tilde\U$ and $\tilde\V$ have orthonormal columns and $\tilde\Si$ is a diagonal matrix.
Notice that the accuracy of the randomized $k$-SVD depends only on the quality of the sketch matrix $\C$.

\subsection{Implementation}

The algorithm is described in Algorithm~\ref{alg:ksvd_prototype} and can be implemented in 5 lines of MATLAB code.
Here $s = \OM(\frac{k}{\epsilon})$ is the size of the sketch.

\vspace{3mm}
\begin{lstlisting}
function [Utilde, Stilde, Vtilde] = ksvdPrototype(A, k, s)
C = CountSketch(A, s);
[Q, R] = qr(C, 0);
[Ubar, Stilde, Vtilde] = svds(Q' * A, k);
Utilde = Q * Ubar;
\end{lstlisting}
\vspace{3mm}
Empirically, using ``$\mathrm{svd(Q' * A,\, 'econ')}$'' followed by discarding the $k+1$ to $s$ components should be faster
than the ``$\mathrm{svds}$'' function in Line 4.

The algorithm has the following properties:
\begin{enumerate}
\item
    The algorithm goes 2 passes through $\A$;
\item
    The algorithm only keeps an $m\times \OM(\frac{k}{\epsilon})$ sketch $\C$ in memory;
\item
    The time cost is $\OM (\nnz (\A) k /\epsilon)$.
\end{enumerate}

\section{Faster Randomized $k$-SVD} \label{sec:ksvd_faster}

The prototype algorithm spends most of its time on solving \eqref{eq:ksvd_prototype};
if \eqref{eq:ksvd_prototype} can be solved more efficiently, the randomized $k$-SVD can be even faster.
The readers may have noticed that \eqref{eq:ksvd_prototype} is the least squares regression (LSR) problem discussed in Section~\ref{sec:lsr:extension1}.
Yes, we can solve \eqref{eq:ksvd_prototype} efficiently by the inexact LSR algorithm presented in the previous section.

\subsection{Theoretical Explanation}

Now we draw a $m\times p$ GaussianProjection+CountSketch matrix $\PP$ and solve this problem:
\begin{equation}\label{eq:ksvd_faster}
\tilde\X \; = \; \argmin_{\rk(\X) \leq k} \| \underbrace{\PP^T \Q_\C}_{p\times s} \underbrace{\X}_{s\times n} - \underbrace{\PP^T \A}_{p\times n} \|_F^2.
\end{equation}
To understand this trick, the readers can retrospect the extension of LSR in Section~\ref{sec:lsr:extension1}.
Let
\[
\PP = \underbrace{\PP_{cs}}_{m\times p_{cs}} \underbrace{\PP_{srht}}_{p_{cs} \times p}
\]
where $p_{cs} = \OM (k/\epsilon + k^2)$
and $p = \OM ( k/\epsilon )$.
The subspace embedding property of RSHT+CountSketch \cite[Theorem 46]{clarkson2013low} implies that
\begin{align*}
&(1+\epsilon)^{-1} \| \Q_\C \tilde\X - \A \|_F^2
\; \leq \; \| \PP^T  (\Q_\C \tilde\X - \A) \|_F^2
\; \leq \; \| \PP^T  (\Q_\C \X^\star - \A) \|_F^2
\; \leq \; (1+\epsilon) \| \Q_\C \X^\star - \A \|_F^2 ,\\
&\Rightarrow \; \| \Q_\C \tilde\X - \A \|_F^2 \; \leq \; (1+\epsilon)^2 \| \Q_\C \X^\star - \A \|_F^2
    \; \leq \; (1+\epsilon)^3 \| \A - \A_k \|_F^2 .
\end{align*}
Here the second inequality follows from the optimality of $\tilde\X$,
and the last inequality follows from the low-rank approximation property of the sketch $\C = \A \S$.
Thus, by solving \eqref{eq:ksvd_faster} we get $k$-SVD up to $1+\OM (\epsilon)$  Frobenius norm relative error.

\subsection{Algorithm Derivation}

The faster randomized $k$-SVD is described in Algorithm~\ref{alg:ksvd_faster}
and derived in the following.
The algorithm solves
\begin{equation}\label{eq:ksvd_faster_c}
\tilde\X \; = \; \argmin_{\rk(\X) \leq k} \| \underbrace{\PP^T \C}_{p\times s} \underbrace{\X}_{s\times n} - \underbrace{\PP^T \A}_{p\times n} \|_F^2
\end{equation}
%which is equivalent to \eqref{eq:ksvd_faster},
to obtain the rank $k$ matrix $\tilde\X \in \RB^{c\times n}$,
and approximates $\A_k$ by
\[
\A_k \approx \C \tilde\X.
\]
Define $\D = \PP^T \C$, $\LL = \PP^T \A$, and let $\Q_\D \R_\D=\D$ be the QR decomposition.
Then \eqref{eq:ksvd_faster_c} becomes
\begin{equation*}%\label{eq:ksvd_faster_c2}
\tilde\X
\; = \; \argmin_{\rk(\X) \leq k} \| \underbrace{\D}_{p\times s} \underbrace{\X}_{s\times n} - \underbrace{\LL}_{p\times n} \|_F^2
\; = \; \underbrace{\R_{\D}^\dag}_{s\times s} \underbrace{(\Q_\D^T \LL)_k}_{s\times n} .
\end{equation*}
%The solution to \eqref{eq:ksvd_faster_c} satisfies $\D \tilde\X = \Q_\D (\Q_\D^T \LL)_k$,
%and thus $\tilde\X = \D^\dag \Q_\D (\Q_\D^T \LL)_k = \R_{\D}^\dag (\Q_\D^T \LL)_k$.
Based on the defined notation, we decompose $\A_k \approx \C \tilde\X$ by
\begin{eqnarray*}
\A_k
& \approx &\C \tilde\X
\; = \; \C \R_{\D}^\dag \underbrace{(\Q_\D^T \LL)_k}_{:= \bar\U \bar{\bf \Sigma} \bar\V^T}
\; = \;\underbrace{ \C \R_{\D}^\dag \bar\U \bar\Si}_{:= \tilde\U \tilde{\bf \Sigma} \hat\V^T} \bar\V^T
\; = \; \tilde\U \tilde{\bf \Sigma} \underbrace{\hat\V^T \bar\V^T }_{:= \tilde\V^T}
\; = \; \underbrace{\tilde\U}_{m\times k} \underbrace{\tilde\Si}_{k\times k} \underbrace{\tilde\V^T }_{k\times n}.
\end{eqnarray*}

\begin{algorithm}[tb]
    \caption{Faster Randomized $k$-SVD Algorithm.}
    \label{alg:ksvd_faster}
\begin{small}
\begin{algorithmic}[1]
    \STATE {\bf Input:} an $m\times n$ matrix $\A$ and the target rank $k$.
    \STATE Set the parameters as $s = \tilde\OM(\frac{k}{\epsilon})$,
        $p_{cs} = s^2 \log^6 \frac{s}{\epsilon} + \frac{s}{\epsilon}$, and $p =\frac{s}{\epsilon} \log \frac{s}{\epsilon}$;
    \STATE Draw a $n\times s$ count sketch matrix $\S$ and perform sketching: $\C = \A \S$;
    \STATE Draw an $m\times p_{cs}$ count sketch matrix $\PP_{cs}$ and an $p_{cs} \times p$ matrix $\PP_{srht}$;
    \STATE Perform Sketching: $\D = \PP_{srht}^T \PP_{cs}^T \C \in \RB^{p \times s}$
            and $\LL = \PP_{srht}^T \PP_{cs}^T \A \in \RB^{p \times n}$;
    %\STATE {\it // Solve the problem \; $\min_{\rk (\X)\leq k} \| \D \X - \LL\|_F^2 \; = \; \min_{\rk (\Z)\leq k} \| \Q_\D \Z - \LL\|_F^2$}
    \STATE QR decomposition: $[\underbrace{\Q_\D}_{p \times s} , \underbrace{\R_\D}_{s\times s}] = qr(\underbrace{\D}_{p \times s})$;
    \STATE $k$-SVD: $[\underbrace{\bar\U}_{s\times k} , \underbrace{\bar\Si}_{k\times k} , \underbrace{\bar\V}_{n\times k}]
            = svds(\underbrace{\Q_\D^T \LL}_{s\times n}, k)$;
    \STATE SVD: $[\underbrace{\tilde\U}_{n\times k} , \underbrace{\tilde\Si}_{k\times k} , \underbrace{\hat\V}_{k\times k}]
            = svd (\underbrace{\C \R_{\D}^\dag \bar\U \bar\Si}_{s\times k})$;
    \STATE $\tilde\V = \underbrace{\bar\V}_{n\times k} \underbrace{\hat\V}_{k\times k}$;
    \RETURN $\tilde\U \tilde\Si \tilde\V^T \approx \A_k$.
\end{algorithmic}
\end{small}
\end{algorithm}

\subsection{Implementation}

The faster randomized $k$-SVD is described in Algorithm~\ref{alg:ksvd_faster} and implemented in 18 lines of MATLAB code.

\vspace{3mm}

\begin{lstlisting}
function [Utilde, Stilde, Vtilde] = ksvdFaster(A, k, s, p1, p2)
n = size(A, 2);
C = CountSketch(A, s);
A = [A, C];
A = A';
sketch = CountSketch(A, p1);
clear A % A (m-by-n) will not be used
sketch = GaussianProjection(sketch, p2);
sketch = sketch';
L = sketch(:, 1:n);
D = sketch(:, n+1:end);
clear sketch % sketch (p2-by-(n+c)) will not be used
[QD, RD] = qr(D, 0);
[Ubar, Sbar, Vbar] = svds(QD' * L, k);
clear L % L (p2-by-n) will not be used
C = C * (pinv(RD) * (Ubar * Sbar));
[Utilde, Stilde, Vhat] = svd(C, 'econ');
Vtilde = Vbar * Vhat;
\end{lstlisting}
\vspace{3mm}
There are a few things to remark:
\begin{enumerate}
\item
    The algorithm goes only two passes through $\A$.
\item
    The algorithm costs time $\OM \big(\nnz (\A) + (m+n) \poly (k/\epsilon) \big)$.
\item
    The parameters should be set as $k < s < p2 < p1 \ll m, n$.
\item
    Line 8 can be removed or replaced by other sketching methods.
\item
    ``A'', ``sketch'', and ``L'' are the most memory expensive variables in the program,
    but fortunately, they are swept only one or two passes.
    If ``A'', ``sketch'', and ``L'' do not fit in memory,
    they should be stored in disk and loaded to memory block-by-block to perform computations.
\item
    Unless both $m$ and $n$ are large enough, this algorithm may be slower than the prototype algorithm.
\end{enumerate}

%%%%%%%%%%%%%%%%%%%%%%%%%%%%%%%%%%%%%%%%%%%%%%%%%%%%%%%%%%%%%%%%%%%%%%%%%%%%%%
%%%%%%%%%%%%%%%%%%%%%%%%%%%%%%%%%%%%%%%%%%%%%%%%%%%%%%%%%%%%%%%%%%%%%%%%%%%%%%
%%%%%%%%%%%%%%%%%%%%%%%%%%%%%%%%%%%%%%%%%%%%%%%%%%%%%%%%%%%%%%%%%%%%%%%%%%%%%%
%%%%%%%%%%%%%%%%%%%%%%%%%%%%%%%%%%%%%%%%%%%%%%%%%%%%%%%%%%%%%%%%%%%%%%%%%%%%%%

\chapter{SPSD Matrix Sketching} \label{sec:spsd}

This chapter considers SPSD matrix $\K \in \RB^{n\times n}$,
which can be a kernel matrix, a social network graph, a Hessian matrix, or a Fisher information matrix.
Our objective is to find a low-rank decomposition $\K \approx \LL \LL^T$.
(Notice that $\LL \LL^T$ is always SPSD, no matter what $\LL$ is.)
If $\K$ is symmetric but not SPSD, it can be approximated by $\K \approx \C \Z \C^T$ where $\Z$ is symmetric but not necessarily SPSD.

\section{Motivations} \label{sec:spsd_motivation}

This section provides three motivation examples to show why we seek to sketch $\K$ by $\K \approx \LL \LL^T$ or $\K \approx \C \Z \C^T$.

\subsection{Forming a Kernel Matrix}

In the kernel approximation problems, we are given
\begin{itemize}
\item
    an $n\times d$ matrix $\X$, whose rows are data points $\x_1 , \cdots , \x_n \in \RB^d$,
\item
    a kernel function, e.g.\ the Gaussian RBF kernel function defined by
    \[
    \kappa (\x_i, \x_j) = \exp \Big(-\frac{1}{2\sigma^{-2}} \|\x_i - \x_j\|_2^2 \Big)
    \]
    where $\sigma > 0$ is the kernel width parameter.
\end{itemize}
The RBF kernel matrix can be computed by the following MATLAB code:

\vspace{3mm}

\begin{lstlisting}
function [K] = rbf(X1, X2, sigma)
K = X1 * X2';
X1_row_sq = sum(X1.^2, 2) / 2;
X2_row_sq = sum(X2.^2, 2) / 2;
K = bsxfun(@minus, K, X1_row_sq);
K = bsxfun(@minus, K, X2_row_sq');
K = K / (sigma^2);
K = exp(K);
\end{lstlisting}
\vspace{3mm}
If $\X_1$ and $\X_2$ are respectively $n_1\times d$ and $n_2\times d$ matrices,
then the output of ``$\mathrm{rbf}$'' is an $n_1\times n_2$ matrix.

Kernel methods requires forming the $n\times n$ kernel matrix $\K$ whose the $(i,j)$-th entry is $\kappa (\x_i , \x_j)$.
The RBF kernel matrix can be computed by the MATLAB function
\begin{lstlisting}
K = rbf(X, X, sigma)
\end{lstlisting}
in $\OM (n^2 d)$ time.

In presence of millions of data points,
it is prohibitive to form such a kernel matrix.
Fortunately, a sketch of $\K$ can be obtained very efficiently.
Let $\S \in \RB^{n\times s}$ be a uniform column selection
matrix\footnote{The local landmark selection is sometimes a better choice.
Do not use random projections, because they inevitably visit every entry of $\K$.}
described in Section~\ref{sec:columnselection},
then $\C = \K \S$ can be obtained in $\OM (n s d)$ time by the following MATLAB code.

\vspace{3mm}

\begin{lstlisting}
function [C] = rbfSketch(X, sigma, s)
n = size(X, 1);
idx = sort(randsample(n, s));
C = rbf(X, X(idx, :), sigma);
\end{lstlisting}
\vspace{3mm}

\subsection{Matrix Inversion} \label{sec:spsd_inversion}

Let $\K$ be an $n\times n$ kernel matrix, $\y$ be an $n$ dimensional vector, and $\alpha$ be a positive constant.
Kernel methods such as the Gaussian process regression (or the equivalent the kernel ridge regression) and the least squares SVM require solving
\[
(\K + \alpha \I_n) \w \; = \; \y
\]
to obtain $\w \in \RB^n$.
The exact solution costs $\OM(n^3)$ time and $\OM(n^2)$ memory.

If we have a rank $l$ approximation $\K \approx \LL \LL^T$,
then $\w$ can be approximately obtained in $\OM (n l^2)$ time and $\OM (n l)$ memory.
Here we need to apply the Sherman-Morrison-Woodbury matrix identity
\[
(\A + \B \C \D)^{-1} \; = \; \A^{-1} - \A^{-1} \B (\C^{-1} + \D \A^{-1} \B)^{-1} \D \A^{-1} .
\]
We expand $(\LL \LL^T + \alpha \I_n )^{-1}$ by the above identity and obtain
\[
(\LL \LL^T + \alpha \I_n )^{-1}
\; = \;
\alpha^{-1} \I_n - \alpha^{-1} \LL (\underbrace{\alpha \I_l + \LL^T \LL}_{l\times l})^{-1} \LL^T ,
\]
and thus
\[
\w \; = \; (\K + \alpha \I_n)^{-1}\y
\; \approx \; \alpha^{-1}  \y - \alpha^{-1}  \LL (\alpha \I_l + \LL^T \LL)^{-1} \LL^T \y .
\]

\vspace{3mm}

The matrix inversion problem not only appears in the kernel methods,
but also in the second order optimization problems.
Newton's method and the so-called natural gradient method require computing $\H^{-1} \g$,
where $\g$ is the gradient and $\H$ is the Hessian matrix or the Fisher information matrix.
Since low-rank matrices are not invertible, the naive low-rank approximation $\H \approx \C \Z \C^T$ does not work.
To make matrix inversion possible, one can use the spectral shifting trick of \cite{wang2014modified}:
fix a small constant $\alpha > 0$, form the low-rank approximation $\H - \alpha \I_n \approx \C \Z \C^T$,
and compute $\H^{-1} \g \approx (\C \Z \C^T + \alpha \I_n )^{-1} \g$.
Besides the low-rank approximation approach, one can approximate $\H$ by a block diagonal matrix or even its diagonal,
because it is easy to invert a diagonal matrix or a block diagonal matrix.

\subsection{Eigenvalue Decomposition}

With the low-rank decomposition $\K \approx \LL \LL^T$ at hand,
we first approximately decompose $\K$ by
\[
\K \; \approx \; \LL \LL^T
\; = \; (\U_\LL \Si_\LL \V_\LL^T) (\U_\LL \Si_\LL \V_\LL^T)^T
\; = \; \U_\LL \Si_\LL^2 \U_\LL^T,
\]
and then discard the $k+1$ to $l$ components in $\U_\LL$ and $\Si_\LL$.
Here $\LL = \U_\LL \Si_\LL \V_\LL^T$ is the SVD of $\LL$, which can be obtained in $\OM (n l^2)$ time and $\OM (n l)$ memory.
In this way, the rank $k$ ($k \leq \rk(\LL)$) eigenvalue decomposition is approximately computed.

\section{Prototype Algorithm} \label{sec:spsd_prototype}

From now on, we will consider how to find the low-rank approximation $\K \approx \LL \LL^T $.
As usual, the simplest approach is to form a sketch $\C = \K \S \in \RB^{n\times s}$ and solve
\begin{equation} \label{eq:spsd:prototype}
\X^\star \; = \; \min_\X \| \K -  \C \X \C^T \|_F^2 \; = \; \C^\dag \K (\C^\dag)^T
\quad \textrm{ or } \quad
\Z^\star \; = \; \min_\Z \| \K -  \Q_\C \Z \Q_\C \|_F^2\; = \; \Q_\C^T \K \Q_\C ,
\end{equation}
where
$\Q_\C$ is the orthonormal bases of $\C$ computed by SVD or QR decomposition.
It is obvious that $\C \X^\star \C = \Q_\C \Z^\star \Q_\C^T$.
In this way, a rank $c$ approximation to $\K$ is obtained.
This approach is first studied by \cite{halko2011ramdom}.
Wang \etal \cite{wang2014modified} showed that if $\C$ contains $s=\OM (k/\epsilon)$ columns of $\K$ chosen by adaptive sampling,
the error bound
\[
\EB \| \K - \Q_\C \Z^\star \Q_\C^T \|_F^2 \; \leq \; (1+\epsilon) \| \K - \K_k\|_F^2
\]
is guaranteed.
Other sketching methods can also be applied, although currently they do not have $1+\epsilon$ error bound.
In the following we implement the prototype algorithm (with the count sketch) in 5 lines of MATLAB code.
Since the algorithm goes only two passes through $\K$,
when $\K$ does not fit in memory, we can store $\K$ in the disk and keep one block of $\K$ in memory at a time.
In this way, $\OM (n s)$ memory is enough.
\vspace{3mm}

\begin{lstlisting}
function [QC, Z] = spsdPrototype(K, s)
n = size(K, 2);
C = CountSketch(K, s);
[QC, ~] = qr(C, 0);
Z = QC' * K * QC;
\end{lstlisting}
\vspace{3mm}
Despite its simplicity, the algorithm has several drawbacks.
\begin{itemize}
\item
    The time cost of this algorithm is $\OM (n s^2 + \nnz(\K) s)$, which can be quadratic in $n$.
\item
    The algorithm must visit every entry of $\K$, which can be a serious drawback when applied to kernel methods.
    It is because computing the kernel matrix $\K$ costs $\OM (n^2 d)$ time, where $d$ is the dimension of the data points.
\end{itemize}
Therefore, we are interested in computing a low-rank approximation in linear time (w.r.t.\ $n$) and avoiding visiting every entry of $\K$.

\section{Faster SPSD Matrix Sketching} \label{sec:spsd_faster}

The readers may have noticed that \eqref{eq:spsd:prototype} is the problem studied in Section~\ref{sec:lsr:extension2}.
We can thus draw a column selection matrix $\PP \in \RB^{n\times p}$ and approximately solve \eqref{eq:spsd:prototype}  by
\begin{equation} \label{eq:spsd:faster}
\tilde\Z \; = \; \min_\Z \| \PP^T (\K -  \Q_\C \Z \Q_\C) \PP \|_F^2
\; = \; \underbrace{(\PP^T \Q_\C)^\dag}_{s\times p} \underbrace{(\PP^T \K \PP)}_{p\times p} \underbrace{( \Q_\C^T \PP)^\dag}_{p\times s} .
\end{equation}
Then we can approximate $\K$ by $\Q_\C \tilde\Z \Q_\C^T$.
We describe the faster SPSD matrix sketching in Algorithm~\ref{alg:spsd_faster}.

\begin{algorithm}[tb]
    \caption{Faster SPSD Matrix Sketching.}
    \label{alg:spsd_faster}
\begin{small}
\begin{algorithmic}[1]
    \STATE {\bf Input:} an $n\times n$ matrix $\K$ and integers $s$ and $p$ ($s \leq p \ll n$).
    \STATE Draw a column selection matrix $\S \in \RB^{n\times s}$;
    \STATE Perform sketching: $\C = \A \S$;
    \STATE QR decomposition: $[\Q_\C , \R_\C] = qr (\C)$;
    \STATE Draw a column selection matrix $\PP \in \RB^{n\times p}$;
    \STATE Compute $\tilde\Z = (\PP^T \Q_\C)^\dag (\PP^T \K \PP) (\Q_\C^T \PP)^\dag$;
    \RETURN $\Q_\C \tilde\Z \Q_\C^T \approx \A$.
\end{algorithmic}
\end{small}
\end{algorithm}

There are a few things to remark.
\begin{itemize}
\item
    Since we are trying to avoid computing every entry of $\K$,
    we should use uniform sampling or local landmark selection to form $\C = \K \S$.
\item
    Let $\PP \in \RB^{n\times p}$ be a leverage score sampling matrix according to the columns of $\C^T$.
    That is, it samples the $i$-th column with probability proportional to $q_i$,
    where $q_i$ is the squared $\ell_2$ norm of the $i$-th row of $\Q_\C$ (for $i = 1$ to $n$).
    When $p = \OM(\sqrt{n} s \epsilon^{-1/2}) $, the following error bounds holds with high probability \cite{wang2015towards}
    \[
     \| \K - \Q_\C \tilde\Z \Q_\C^T \|_F^2 \; \leq \; (1+\epsilon) \min_{\Z} \| \K - \Q_\C \Z \Q_\C^T \|_F^2 .
    \]
\item
    Let $\S$ be a uniform sampling matrix and $\PP$ be a leverage score sampling matrix.
    The algorithm visits only $ns + p^2 = \OM (n)$ entries of $\K$.
    The overall time and memory costs are linear in $n$.
\item
    Assume $\S$ is a column selection matrix.
    Let the sketch $\C = \K \S$ contains the columns of $\K$ indexed by $\SM \subset [n]$,
    and the columns selected by $\PP$ are indexed by $\PM \subset [n]$.
    Empirically, enforcing $\SM \subset \PM$ significantly improves the approximation quality.
\item
    Empirically, letting $p$ be several times larger than $s$, e.g.\ $p = 4s$, is sufficient for a high quality.
\end{itemize}
The algorithm can be implemented in 12 lines of MATLAB code.

\vspace{3mm}

\begin{lstlisting}
function [QC, Z] = spsdFaster(K, s)
p = 4 * s; % can be tuned
n = size(K, 2);
S = sort(randsample(n, s)); % uniform sampling
C = K(:, S);
[QC, ~] = qr(C, 0);
q = sum(QC.^2, 2); % the sampling probability
q = q / sum(q);
P = randsample(n, p,true, q); % leverage score sampling
P = unique([P; S]); % enforce P to contain S
PQCinv = pinv(QC(P, :));
Z = PQCinv * K(P, P) * PQCinv';
\end{lstlisting}

\vspace{3mm}

The above implementation assumes that $\K$ is a given matrix.
In the kernel approximation problems, we are only given a $n\times d$ matrix $\X$, whose rows are data points,
and a kernel function, e.g.\ the RBF kernel with width parameter $\sigma$.
We should implement the faster SPSD sketching algorithm in the following way.

\vspace{3mm}

\begin{lstlisting}
function [QC, Z] = spsdFaster(X, sigma, s)
p = 4 * s; % can be tuned
n = size(X, 1);
S = sort(randsample(n, s)); % uniform sampling
C = rbf(X, X(S, :), sigma);
[QC, ~] = qr(C, 0);
q = sum(QC.^2, 2); % the sampling probability
q = q / sum(q);
P = randsample(n, p,true, q);
P = unique([P; S]); % enforce P contains S
PQCinv = pinv(QC(P, :));
Ksub = rbf(X(P, :), X(P, :), sigma);
Z = PQCinv * Ksub * PQCinv';
\end{lstlisting}
\vspace{3mm}
The above implementation avoids computing the whole kernel matrix,
and is thus highly efficient when applied to kernel methods.

\section{The Nystr\"om Method} \label{sec:nystrom}

\begin{figure}[!ht]
\begin{center}
\centering
\includegraphics[width=0.85\textwidth]{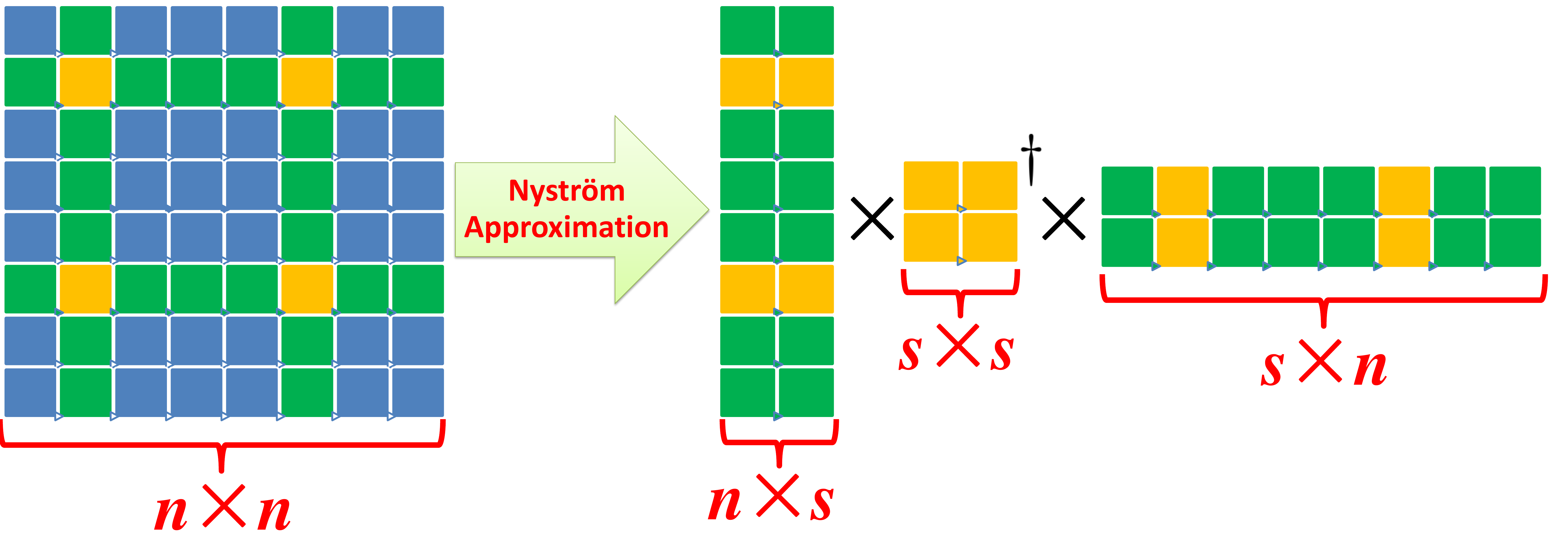}
\end{center}
   \caption{The illustration of the Nystr\"om method.}
\label{fig:nystrom}
\end{figure}

Let $\S$ be an $n\times s$ column selection matrix and $\C = \K \S \in \RB^{n\times s}$ be a sketch of $\K$.
Recall the model \eqref{eq:spsd:faster} proposed in the previous section.
It is easy to verify that $\Q_\C \tilde\Z \Q_\C^T = \C \tilde\X \C^T$, where $\tilde\X$ is defined by
\[
\tilde\X \; = \; \min_\X \| \PP^T (\K -  \C \X \C) \PP \|_F^2
\; = \; \underbrace{(\PP^T \C)^\dag}_{s\times p} \underbrace{(\PP^T \K \PP)}_{p\times p} \underbrace{( \C^T \PP)^\dag}_{p\times s} .
\]
One can simply set $\PP = \S \in \RB^{n\times s}$ and let $\W = \S^T \C = \S^T \K \S$.
Then the solution $\tilde\X$ becomes
\[
\tilde\X \; = \; (\S^T \C)^\dag (\S^T \K \S) (\C^T \S)^\dag
\; = \; \W^\dag \W \W^\dag \; = \; \W^\dag .
\]
The low-rank approximation
\[
\K \approx \C \W^\dag \C^T
\]
is called the Nystr\"om method \cite{nystrom1930praktische,williams2001using}.
The Nystr\"om method is perhaps the most extensively used kernel approximation approach in the literature.
See Figure \ref{fig:nystrom} for the illustration of the Nystr\"om method.

%---------------------------------Figure---------------------------------%
%---------------------------------Figure---------------------------------%

There are a few things to remark:
\begin{itemize}
\item
    The Nystr\"om is highly efficient.
    When applied to speedup kernel methods, the scalability can be as large as $n = 10^6$.
\item
    The Nystr\"om method is a rough approximation to $\K$ and is well known to be of low accuracy.
    If a moderately high accuracy is required, one had better use the method in the previous section.
\item
    The $s\times s$ matrix $\W$ is usually ill-conditioned, and thus the Moore-Penrose inverse can be numerically instable.
    (It is because the bottom singular values of $\W$ blow up during the Moore-Penrose inverse.)
    A very effective heuristic is to drop the bottom singular values of $\W$:
    set a parameter $k < s$, e.g.\ $k = \lceil 0.8 s\rceil$, and approximate $\K$ by $\C (\W_k)^\dag \C^T$.
\item
    There are many choices of the sampling matrix $\S$. See \cite{gittens2013revisiting} for more discussions.
\end{itemize}
The Nystr\"om method can be implemented in $11$ lines of MATLAB code.
The output of the algorithm is $\LL \in \RB^{n\times k}$, where $\LL \LL^T$ is the Nystr\"om approximation to
$\K$.\footnote{Let $\W_k = \U_{\W,k} \Lam_{\W,k} \U_{\W,k}^T$ be the $k$-eigenvalue decomposition of $\W$
and set $\LL = \C \U_{\W,k} \Lam_{\W,k}^{-1} \in \RB^{n\times k}$.}

\vspace{3mm}

\begin{lstlisting}
function [L] = Nystrom(X, sigma, s)
k = ceil(0.8 * s); % can be tuned
n = size(X, 1);
S = sort(randsample(n, s)); % uniform sampling
C = rbf(X, X(S, :), sigma); % C = K(:, S)
W = C(S, :);
[UW, SW, ~] = svd(W);
SW = diag(SW);
SW = 1 ./ sqrt(SW(1:k));
UW = bsxfun(@times, UW(:, 1:k), SW');
L = C * UW; % K is approximated by L * L'
\end{lstlisting}
\vspace{3mm}
Here we use the RBF kernel function implemented in Section~\ref{sec:spsd_motivation}.
Line 8 sets $k = \lceil 0.8 c\rceil$, which can be better tuned to enhance numerical stability.
Notice that $k$ should not be set too small, otherwise the accuracy would be affected.

\section{More Efficient Extensions} \label{sec:spsd_moreefficient}

Several SPSD matrix approximation methods has been proposed recently,
and they are more scalable than the Nystr\"om method in certain applications.
This section briefly describes some of these methods.

\subsection{Memory Efficient Kernel Approximation (MEKA)}

MEKA \cite{si2014memory}
exploits the block structure of kernel matrices and is more memory efficient than the Nystr\"om method.
MEKA first partitions the data $\x_1 , \cdots ,\x_n$ into $b$ groups (e.g.\ by inexact $k$means clustering),
accordingly, the kernel matrix $\K$ has $b\times b$ blocks:
\[
\K \; = \; \left[
             \begin{array}{ccc}
               \K_{[1,1]} & \cdots & \K_{[1,b]}  \\
               \vdots & \ddots & \vdots \\
               \K_{[b,1]} & \cdots & \K_{[b,b]}  \\
             \end{array}
           \right]
 \; = \; \left[
             \begin{array}{c}
               \K_{[1:]}   \\
               \vdots \\
               \K_{[b:]}   \\
             \end{array}
           \right].
\]
Then MEKA approximately computes the top left singular vectors of $ \K_{[1:]} , \cdots ,  \K_{[b:]}$, denote $ \U_{[1]}$, $\cdots, \U_{[b]}$, respectively.
For each $(i, j) \in [b]\times [b]$,
MEKA finds a very small-scale matrix $\Z_{[i,j]}$ by solving
\[
 \Z_{[i,j]}
\; = \; \argmin_\Z \big\| \K_{[i,j]} - \U_{[i]} \Z_{[i,j]} \U_{[j]}^T \big\|_F^2 .
\]
This can be done efficiently using the approach in Section~\ref{sec:lsr:extension2}.
Finally, the low-rank approximation is
\[
\K \; \approx \; \left[
                   \begin{array}{ccc}
                     \U_{[1]} &   & \0 \\
                      & \ddots &  \\
                     \0 &  & \U_{[b]} \\
                   \end{array}
                 \right]
                 \left[
             \begin{array}{ccc}
               \Z_{[1,1]} & \cdots & \Z_{[1,b]}  \\
               \vdots & \ddots & \vdots \\
               \Z_{[b,1]} & \cdots & \Z_{[b,b]}  \\
             \end{array}
           \right]
                 \left[
                   \begin{array}{ccc}
                     \U_{[1]} &   & \0 \\
                      & \ddots &  \\
                     \0 &  & \U_{[b]} \\
                   \end{array}
                 \right]^T
           \; = \; \U \Z \U^T.
\]
Since $\Z$ and $\U_{[1]}, \cdots , \U_{[b]}$ are small-scale matrices, MEKA is thus very memory efficient.
There are several things to remark:
\begin{itemize}
\item
    MEKA can be used to speedup Gaussian process regression and least squares SVM.
    However, MEKA can be hardly applied to speedup $k$-eigenvalue decomposition,
    because it requires the $k$-SVD of $\U \Z^{1/2}$, which destroys the sparsity and significantly increases memory cost.
\item
    Indiscreet implementation, e.g.\ the implementation provided by \cite{si2014memory}, can make MEKA numerically unstable,
    as was reported by \cite{wang2014modified,wang2015structured}.
    The readers had better to follow the stabler implementation in \cite{wang2015structured}.
\end{itemize}

\subsection{Structured Kernel Interpolation (SKI)}

SKI \cite{wilson2015kernel} is a memory efficient extension of the Nystr\"om method.
Let $\S$ be a column selection matrix, $\C = \K \S$, and $\W = \S^T \C = \S^T \K \S$.
The Nystr\"om method approximates $\K$ by $\C \W^\dag \C^T$.
SKI further approximates each row of $\C$ by a convex combination of two rows of $\W$
and obtain $\C \approx \X \W$.
Notice that each row of $\X$ has only two nonzero entries, which makes $\X$ extremely sparse.
In this way, $\K$ is approximated by
\[
\K \; \approx \;
\C \W^\dag \C
\; \approx \;
(\X \W) \W^\dag (\X \W)^T
\; = \; \X \W \X^T.
\]
Much accuracy is lost in the second approximation,
so SKI is much less accurate than the Nystr\"om method.
For the same reason as MEKA, there is no point in applying SKI to speedup $k$-eigenvalue decomposition of $\K$.

\section{Extension to Rectangular Matrices: CUR Matrix Decomposition} \label{sec:spsd_cur}

This section considers the problem of sketching any rectangular matrix $\A$ by the CUR matrix decomposition~\cite{mahoney2011ramdomized}.
The CUR matrix decomposition is an extension of the previously discussed SPSD matrix sketching methods.

\subsection{Motivation} \label{sec:spsd_cur:motivation}
Suppose we are given $n$ training data $\x_1 , \cdots , \x_n \in \RB^d$,
$m$ test data $\x_1', \cdots , \x_m' \in \RB^d$, and a kernel function $\kappa (\cdot , \cdot)$.
In their generalization (test) stage,
kernel methods such as GPR and KPCA form an $m\times n$ matrix $\K_*$,
where $(\K_*)_{i j} = \kappa (\x_i' , \x_j)$,
and apply $\K_*$ to some vectors or matrices.
Notice that it takes $\OM (m n d)$ time to form $\K_*$ and $\OM (m n p)$ time to multiply $\K_*$ by an $n\times p$ matrix.
If $m$ is as large as $n$, the generalization stage of such kernel methods can be very expensive.
Fortunately, with the help of the CUR matrix decomposition,
the generalization stage of GPR or KPCA merely costs time linear in $m+n$.

\subsection{Prototype CUR Decomposition}

Suppose we are given an arbitrary $m\times n$ rectangular matrix $\A$, which can be the aforementioned $\K_*$.
We sample $c$ columns of $\A$ to form $\C = \A \S_\C \in \RB^{m\times c}$,
sample $r$ rows of $\A$ to form $\R  = \A \S_\R \in \RB^{r\times n}$, and compute the intersection matrix $\U^\star \in \RB^{c\times r}$ by solving
\begin{equation} \label{eq:cur_prototype}
\U^\star \; = \;
\argmin_{\U} \| \underbrace{\A}_{m\times n} - \underbrace{\C}_{m\times c} \underbrace{\U}_{c\times r} \underbrace{\R}_{r\times n} \|_F^2
\; = \; \C^\dag \A \R^\dag  .
\end{equation}
The approximation $\A \approx \C \U^\star \R$ is well known as the CUR decomposition~\cite{mahoney2011ramdomized}.
This formulation bears a strong resemblance with the prototype SPSD matrix sketching method in \eqref{eq:spsd:prototype}.

The prototype CUR decomposition is not very useful because
(1) its time cost is $\OM(mn \cdot \min\{c, r\})$ and (2) it visits every entry of $\A$.

\subsection{Faster CUR Decomposition} \label{sec:spsd_cur:faster}

Analogous to the SPSD matrix sketching,
we can compute $\U^\star$ approximately and significantly more efficiently.
Let $\PP_\C \in \RB^{m\times p_c}$ and $\PP_\R \in \RB^{n\times p_r}$ be some column selection matrices.
Then we solve this problem in stead of \eqref{eq:cur_prototype}:
\begin{equation} \label{eq:cur_faster}
\tilde\U \; = \;
\argmin_{\U} \| \underbrace{\PP_\C^T \A \PP_\R}_{p_c \times p_r} -
\underbrace{\PP_\C^T \C}_{p_c \times c} \underbrace{\U}_{c\times r} \underbrace{\R \PP_\R}_{r\times p_r} \|_F^2
\; = \; (\PP_\C^T \C)^\dag (\PP_\C^T \A \PP_\R) (\R \PP_\R)^\dag  .
\end{equation}
The faster CUR decomposition is very similar to the faster SPSD matrix sketching method in Section~\ref{sec:spsd_faster}.
The faster CUR decomposition has the following properties:
\begin{itemize}
\item
    It visits only $m c + n r + p_c p_r$ entries of $\A$, which is linear in $m+n$.
    This is particularly useful when applied to kernel methods,
    because it avoids forming the whole kernel matrix.
\item
    The overall time and memory costs are linear in $m+n$.
\item
    If $\PP_\C$ is the leverage score sampling matrix corresponding to the columns of $\C^T$
    and $\PP_\R$ is the leverage score sampling matrix corresponding to the columns of $\R$,
    then $\tilde\U$ is a very high quality approximation to $\U^\star$ \cite{wang2015towards}:
    \[
    \| \A - \C \tilde\U \R \|_F^2 \; \leq \; (1+\epsilon) \min_\U \|\A - \C \U \R\|_F^2
    \]
    holds with high probability.
\end{itemize}
Empirically speaking,
setting $\PP_\C$ and $\PP_\R$ be uniform sampling matrices works nearly as well as leverage score sampling matrices,
and setting $p_c = p_r = \OM(c+r)$ suffices for a high approximation quality.
If $\A$ is a full-observed matrix, the CUR matrix decomposition can be computed by the following MATLAB code.

\vspace{3mm}

\begin{lstlisting}
function [C, U, R] = curFaster(A, c, r)
pc = 2 * (r + c); % can be tuned
pr = 2 * (r + c); % can be tuned
[m, n] = size(A);
SC = sort(randsample(n, c));
SR = sort(randsample(m, r));
C = A(:, SC);
R = A(SR, :);
PC = sort(randsample(m, pc));
PR = sort(randsample(n, pr));
PC = unique([PC; SR]); % enforce PC to contain SR
PR = unique([PR; SC]); % enforce PR to contain SC
U = pinv(C(PC, :)) * A(PC, PR) * pinv(R(:, PR));
\end{lstlisting}
\vspace{3mm}

Let's consider the kernel approximation problem in Section~\ref{sec:spsd_cur:motivation}.
Let $\X_{\textrm{train}} \in \RB^{n\times d}$ be the training data
and $\X_{\textrm{test}} \in \RB^{m\times d}$ be the test data.
We use the RBF kernel with kernel width parameter $\sigma$.
The $m\times n$ matrix $\K_*$ can be approximated by $\tilde\K_* = \C \U \R$,
which is the output of the following MATLAB procedure.

\vspace{3mm}

\begin{lstlisting}
function [C, U, R] = curFasterKernel(Xtest, Xtrain, sigma, c, r)
pc = 2 * (r + c); % can be tuned
pr = 2 * (r + c); % can be tuned
m = size(Xtest, 1);
n = size(Xtrain, 1);
SC = sort(randsample(n, c));
SR = sort(randsample(m, r));
C = rbf(Xtest, Xtrain(SC, :), sigma);
R = rbf(Xtest(SR, :), Xtrain, sigma);
PC = sort(randsample(m, pc));
PR = sort(randsample(n, pr));
PC = unique([PC; SR]); % enforce PC to contain SR
PR = unique([PR; SC]); % enforce PR to contain SC
Kblock = rbf(Xtest(PC, :), Xtrain(PR, :), sigma);
U = pinv(C(PC, :)) * Kblock * pinv(R(:, PR));
\end{lstlisting}
\vspace{3mm}
The time cost of this procedure is linear in $m+n$,
and $\tilde\K_* = \C \U \R$ can be applied to $n$ dimensional vector in $\OM \big( nr + m c)$ time.
In this way, the generalization of GPR and KPCA can be efficient.

\section{Applications} \label{sec:spsd_application}

This section provides the implementations of kernel PCA, spectral clustering, Gaussian process regression,
all sped-up by randomized algorithms.

\subsection{Kernel Principal Component Analysis (KPCA)}

Suppose we are given
\begin{itemize}
\item
    $n$ training data $\x_1 , \cdots , \x_n \in \RB^d$,
\item
    $m$ test data $\x_1', \cdots , \x_m' \in \RB^d$, ($\x_i'$ is not the transpose $\x_i^T$),
\item
    a kernel function $\kappa (\cdot , \cdot )$, e.g.\ the RBF kernel function,
\item
    a target rank $k$ ($\ll n, d$).
\end{itemize}
The goal of KPCA is to extract $k$ features of each training datum and each test datum,
which may be used in clustering or classification.
The standard KPCA consists of the following steps:
\begin{enumerate}
\item
    Training
    \begin{enumerate}
    \item
        Form the $n\times n$ kernel matrix $\K$ of the training data, whose the $(i,j)$-th entry is $\kappa (\x_i , \x_j)$;
    \item
        Compute the $k$-eigenvalue decomposition $\K_k = \U_k \Lam_k \U_k^T$;
    \item
        Form the $n\times k$ matrix $\U_k \Lam_k^{1/2}$, whose the $i$-th row is the feature of $\x_i$;
    \end{enumerate}
\item
    Generalization (test)
    \begin{enumerate}
    \item
        Form the $m\times n$ kernel matrix $\K_*$ whose the $(i,j)$-th entry is $\kappa (\x_i', \x_j)$;
    \item
        Form the $m\times k$ matrix $\K_* \U_k \Lam_k^{-1/2}$, whose the $i$-th row is the feature of $\x_i'$.
    \end{enumerate}
\end{enumerate}
The most time and memory expensive step in training is the $k$-eigenvalue decomposition of $\K$,
which can be sped-up by the sketching techniques discussed in this section.
Empirically, the faster SPSD matrix sketching in Section~\ref{sec:spsd_faster} is much more accurate than the Nystr\"om method in Section~\ref{sec:nystrom},
and their time and memory costs are all linear in $n$.
Thus the faster SPSD matrix sketching can be better choice.
KPCA can be approximately solved by several lines of MATLAB code.

\vspace{3mm}

\begin{lstlisting}
function [U, lambda, featuretrain] = kpcaTrain(Xtrain, sigma, k)
s = k * 10; % can be tuned
[QC, Z] = spsdFaster(Xtrain, sigma, s); % QC has orthogonal columns
clear Xtrain
[UZ, SZ, ~] = svd(Z);
U = QC * UZ(:, 1:k); % U contains the top k eigenvectors
lambda = diag(SZ);
lambda = lambda(1:k); % lambda is the vector containing the top k eigenvalues
featuretrain = bsxfun(@times, U, (sqrt(lambda))');
end
\end{lstlisting}
\vspace{3mm}

\begin{lstlisting}
function [featuretest] = kpcaTest(Xtrain, Xtest, sigma, U, lambda)
Ktest = rbf(Xtest, Xtrain, sigma);
U = bsxfun(@times, U, (1 ./ sqrt(lambda))');
featuretest = Ktest * U;
end
\end{lstlisting}
\vspace{3mm}
In the function ``$\mathrm{kpcaTrain}$'',
the input variable ``$\mathrm{Xtrain}$'' has $n$ rows, each of which corresponds to a training datum.
The rows of the output ``$\mathrm{featuretrain}$'' and ``$\mathrm{featuretest}$'' are the features extracted by KPCA,
and the features can be used to perform classification.
For example, suppose each datum $\x_i$ is associated with a label $y_i$, and let $\y = [y_1 , \cdots , y_n]^T \in \RB^n$.
We can use $k$-nearest-neighbor
\begin{lstlisting}
[ytest] = knnclassify(featuretest, featuretrain, y)
\end{lstlisting}
to predict the labels of the test data.

When the number of test data $m$ is large,
the function ``$\textrm{kpcaTest}$'' is costly.
The users should apply the CUR decomposition in Section~\ref{sec:spsd_cur:faster} to speedup computation.
\vspace{3mm}

\begin{lstlisting}
function [featuretest] = kpcaTestCUR(Xtrain, Xtest, sigma, U, lambda)
c = max(100, ceil(size(Xtrain, 1) / 20)); % can be tuned
r = max(100, ceil(size(Xtest, 1) / 20)); % can be tuned
[C, Utilde, R] = curFasterKernel(Xtest, Xtrain, sigma, c, r);
U = bsxfun(@times, U, (1 ./ sqrt(lambda))');
featuretest = C * (Utilde * (R * U));
end
\end{lstlisting}
\vspace{3mm}

\subsection{Spectral Clustering}

Spectral clustering is one of the most popular clustering algorithms.
Suppose we are given
\begin{itemize}
\item
    $n$ data points $\x_1 , \cdots ,\x_n \in \RB^d$,
\item
    a kernel function $\kappa (\cdot , \cdot)$,
\item
    $k$: the number of classes.
\end{itemize}
Spectral clustering performs the following operations:
\begin{enumerate}
\item
    Form an $n\times n$ kernel matrix $\K$,
    where big $k_{i j}$ indicates $\x_i$ and $\x_j$ are similar;
\item
    Form the degree matrix $\D$ with $d_{i i} = \sum_{j} k_{i j}$ and $d_{i j} = 0$ for all $i\neq j$;
\item
    Compute the normalized graph Laplacian $\G = \D^{-1/2} \K \D^{-1/2} \in \RB^{n\times n}$;
\item
    Compute the top $k$ eigenvectors of $\G$, denote $\U \in \RB^{n\times k}$, and normalize the rows of $\U$;
\item
    Apply $k$means clustering on the rows of $\V$ to obtain the class labels.
\end{enumerate}
The first step costs $\OM (n^2 d)$ time and the fourth step costs $\OM (n^2 k)$ times,
which limit the scalability of spectral clustering.
Fowlkes \etal \cite{fowlkes2004spectral} proposed to apply the Nystr\"om method to make spectral clustering more scalable
by avoiding forming the whole kernel matrix and speeding-up the $k$-eigenvalue decomposition.
Empirically, the algorithm in Section~\ref{sec:spsd_faster} is more accurate than the Nystr\"om method in Section~\ref{sec:nystrom},
and they both runs in linear time.
Spectral clustering with the randomized algorithm in Section~\ref{sec:spsd_faster} can be implemented in 16 lines of MATLAB code.

\vspace{3mm}

\begin{lstlisting}
function [labels] = SpectralClusteringFaster(X, sigma, k)
s = k * 10; % can be tuned
n = size(X, 1);
[QC, Z] = spsdFaster(X, sigma, s); % K is approximated by QC * Z * QC'
[UZ, SZ, ~] = svd(Z);
SZ = sqrt(diag(SZ));
UZ = bsxfun(@times, UZ, SZ'); % now Z = UZ * UZ'
L = QC * UZ; % now K is approximated by L * L'
d = ones(n, 1);
d = L * (L' * d); % diagonal of the degree matrix D
d = 1 ./ sqrt(d);
L = bsxfun(@times, L, d); % now G is approximated by L*L'
[U, ~, ~] = svd(L, 'econ');
U = U(:, 1:k);
U = normr(U); % normalize the rows of U
labels = kmeans(U, k, 'Replicates', 3);
\end{lstlisting}
\vspace{3mm}
When the scale of data is too large for the faster SPSD matrix sketching algorithm in Section~\ref{sec:spsd_faster},
one can instead use the more efficient Nystr\"om method in Section~\ref{sec:nystrom}:
simply replace Lines 4 to 8 by

\begin{lstlisting}
L = Nystrom(X, sigma, s);
\end{lstlisting}

\subsection{Gaussian Process Regression (GPR)} \label{sec:spsd_gpr}

The Gaussian process regression (GPR) is one of the most popular machine learning methods.
GPR is the foundation of Bayesian optimization
and has important applications such as automatically tuning the hyper-parameters of deep neural networks.
Suppose we are given
\begin{itemize}
\item
    $n$ training data $\x_1 , \cdots , \x_n \in \R^d$,
\item
    labels $\y = [y_1 , \cdots y_n]^T \in \RB^n$ of the training data,
\item
    $m$ test data $\x_1', \cdots , \x_m' \in \RB^d$, ($\x_i'$ is not the transpose $\x_i^T$),
\item
    and a kernel function $\kappa (\cdot , \cdot)$, e.g.\ the RBF kernel with kernel width parameter $\sigma$.
\end{itemize}

\vspace{3mm}

\noindent
{\bf Training.}
In the training stage, GPR requires forming the $n\times n$ kernel matrix $\K$ where $k_{i j}= \kappa (\x_i , \x_j) $
and computing the model
\[
\w \; = \; (\K + \alpha \I_n)^{-1}\y .
\]
Here $\alpha$ is a tuning parameter that indicates the noise intensity in the labels $\y$.
It takes $\OM (n^2 d)$ time to form the kernel matrix and $\OM (n^3)$ time to compute the matrix inversion.
To make the training efficient, we can first sketch the SPSD matrix $\K$ to obtain $\K \approx \LL \LL^T$
and then apply the technique in Section~\ref{sec:spsd_inversion} to obtain $\w$.
Empirically, when applied to speedup GPR, the algorithms discussed in Section~\ref{sec:spsd_faster} and Section~\ref{sec:nystrom}
has similar accuracy, thus we choose to use the Nystr\"om method which is more efficient.

The training GPR with the Nystr\"om approximation can be implemented in the following MATLAB code.
The time cost is $\OM (n l^2 + n l d)$ and the space cost is $\OM (n l + n d)$.

\vspace{3mm}

\begin{lstlisting}
function [w] = gprTrain(Xtrain, ytrain, sigma, alpha)
l = 100; % can be tuned
L = Nystrom(Xtrain, sigma, l); % K is approximated by L * L'
l = size(L, 2);
w = L' * ytrain;
w = (alpha * eye(l) + L' * L) \ w;
w = ytrain - L * w;
w = w / alpha;
end
\end{lstlisting}
\vspace{3mm}
The input ``$\mathrm{sigma}$'' is the kernel width parameter and
``$\mathrm{alpha}$'' indicates the noise intensity in the observation.

\vspace{5mm}

\noindent
{\bf Generalization (test).}
After obtaining the trained model $\w \in \RB^n$,
GPR can predict the unknown labels of the $m$ test data $\x_1', \cdots , \x_m' \in \RB^d$.
GPR forms an $m\times n$ kernel matrix $\K_*$ whose the $(i,j)$-th entry is $\kappa (\x_i', \x_j)$
and compute $\y_* = \K_* \w \in \RB^{m}$.
The $i$-th entry in $\y_*$ is the predictive label of $\x_i'$.
The generalization can be implemented in four lines of MATLAB code.

\vspace{3mm}

\begin{lstlisting}
function [ytest] = gprTest(Xtrain, Xtest, sigma, w)
Ktest = rbf(Xtest, Xtrain, sigma);
ytest = Ktest * w;
end
\end{lstlisting}
\vspace{3mm}
It costs $\OM (m n d)$ time to compute $\K_*$ and $\OM (m n)$ time to apply $\K_*$ to $\w$.
If $m$ is small, the generalization stage can be performed straightforwardly.
However, if $m$ is as large as $n$, the time cost will be quadratic in $n$,
and the user should apply the CUR decomposition in Section~\ref{sec:spsd_cur:faster} to speedup computation.

\vspace{3mm}

\begin{lstlisting}
function [ytest] = gprTestCUR(Xtrain, Xtest, sigma, w)
c = max(100, ceil(size(Xtrain, 1) / 20)); % can be tuned
r = max(100, ceil(size(Xtest, 1) / 20)); % can be tuned
[C, Utilde, R] = curFasterKernel(Xtest, Xtrain, sigma, c, r);
ytest = C * (Utilde * (R * w));
end
\end{lstlisting}

%%%%%%%%%%%%%%%%%%%%%%%%%%%%%%%%%%%%%%%%%%%%%%%%%%%%%%%%%%%%%%%%%%%%%%%%%%%%%%
%%%%%%%%%%%%%%%%%%%%%%%%%%%%%%%%%%%%%%%%%%%%%%%%%%%%%%%%%%%%%%%%%%%%%%%%%%%%%%
%%%%%%%%%%%%%%%%%%%%%%%%%%%%%%%%%%%%%%%%%%%%%%%%%%%%%%%%%%%%%%%%%%%%%%%%%%%%%%
%%%%%%%%%%%%%%%%%%%%%%%%%%%%%%%%%%%%%%%%%%%%%%%%%%%%%%%%%%%%%%%%%%%%%%%%%%%%%%

\appendix

\chapter{Several Facts of Matrix Algebra}

This chapter lists some facts that has been applied in this paper.

\begin{fact}
The matrices $\Q_1 \in \RB^{m\times n}$ and $\Q_{n \times p}$ ($m \geq n \geq p$) have orthonormal columns.
Then the matrix $\Q = \Q_1 \Q_2$ has orthonormal columns.
\end{fact}

%\begin{proof}
%The matrix $\Q$ has orthonormal columns because
%\[
%\Q^T \Q = \Q_2^T (\Q_1^T \Q_1)  \Q_2 = \Q_2^T \I_n \Q_2 = \I_p .
%\]
%\end{proof}

\begin{fact}
Let $\A$ be any $m\times n$ and rank $\rho$ matrix.
Then
$
\A \A^\dag \B = \U_\A \U_\A^T \B = \A \X^\star = \U_\A \Z^\star,
$
where
\[
\X^\star \; = \; \argmin_\X \| \B - \A \X \|_F^2 , \qquad \textrm{ and } \qquad
\Z^\star \; = \; \argmin_\Z \| \B - \U_\A \Z \|_F^2 .
\]
This is the reason why $\A \A^\dag \B$ and $\U_\A \U_\A^T \B $ are called the projection of $\B$ onto the column space of $\A$.
\end{fact}

\begin{fact} \cite[Lemma 44]{woodruff2014sketching}
The matrices $\Q \in \RB^{m\times s}$ ($m\geq s$) has orthonormal columns.
The solution to
\[
\argmin_{\rk (\X) \leq k} \| \A - \Q \X \|_F^2
\]
is $\X^\star = (\Q^T \A )_k$, where $(\Q^T \A )_k$ denotes the closest rank $k$ approximation to $\Q^T \A$.
\end{fact}

\begin{fact}
Let $\A^\dag$ be the Moore-Penrose inverse of $\A$.
Then $\A \A^\dag \A = \A$ and $\A^\dag \A \A^\dag = \A^\dag$.
\end{fact}

\begin{fact}
Let $\A$ be an $m\times n$ ($m\geq n$) matrix and $\A = \Q_\A \R_\A$ be the QR decomposition of $\A$.
Then
\[
\underbrace{\A^\dag}_{n\times m} \; = \; \underbrace{\R_\A^\dag}_{n\times n} \underbrace{\Q_\A^T}_{n\times m} .
\]
\end{fact}

\begin{fact}
Let $\C$ be a full-rank matrix with more rows than columns.
Let $\C = \Q_\C \R_\C$ be the QR decomposition and $\C = \U_\C \Si_\C \V_\C$ be the condensed SVD.
Then the leverage scores of $\C$, $\Q_\C$, $\U_\C$ are the same.
\end{fact}

%\begin{fact}
%Let $\C$ be a matrix with more rows than columns.
%Let $\Q_\C$ be the orthonormal bases of $\C$ computed by the rank revealing QR decomposition
%and
%\[
%\X^\star = \min_\X \| \C \X  - \A \|_F^2  \qquad \textrm{ and } \qquad
%\Z^\star = \min_\Z \| \Q_\C \Z  - \A \|_F^2 .
%\]
%Then $\C$
%\end{fact}

%%%%%%%%%%%%%%%%%%%%%%%%%%%%%%%%%%%%%%%%%%%%%%%%%%%%%%%%%%%%%%%%%%%%%%%%%%%%%%
%%%%%%%%%%%%%%%%%%%%%%%%%%%%%%%%%%%%%%%%%%%%%%%%%%%%%%%%%%%%%%%%%%%%%%%%%%%%%%
%%%%%%%%%%%%%%%%%%%%%%%%%%%%%%%%%%%%%%%%%%%%%%%%%%%%%%%%%%%%%%%%%%%%%%%%%%%%%%
%%%%%%%%%%%%%%%%%%%%%%%%%%%%%%%%%%%%%%%%%%%%%%%%%%%%%%%%%%%%%%%%%%%%%%%%%%%%%%

\chapter{Notes and Further Reading}

{\bf The $\ell_p$ Regression Problems.}
Chapter~\ref{sec:lsr} has applied the sketching methods to solve the $\ell_2$ norm regression problem more efficiently.
The more general $\ell_p$ regression problems have also been studied in the
literature~\cite{clarkson2005subgradient,clarkson2013fast,meng2013low,cohen2014lp}.
Especially, the $\ell_1$ is of great interest because it demonstrate strong robustness to noise.
Currently the strongest result is the $\ell_p$ row sampling by Lewis weights \cite{cohen2014lp}.

{\bf Distributed SVD.}
In the distributed model, each machine holds a subset of columns of $\A$, and the system outputs the top singular values and singular vectors.
In this model, the communication cost should also be considered, as well as the time and memory costs.
The seminal work \cite{feldman2013turning} proposed to build a coreset to capture the properties of $\A$,
which facilitates low computation and communication costs.
Later on, several algorithms with stronger error bound and lower communication and computation costs have been proposed.
Currently, the state of the art is \cite{boutsidis2015communication}.

{\bf Random Feature for Kernel Methods.}
Chapter~\ref{sec:spsd} has introduced the sketching methods for kernel methods.
A parallel line of work is the random feature methods \cite{rahimi2007random}
which also form low-rank approximations to kernel matrices.
Section~6.5.3 of \cite{tropp2015introduction} offers simple and elegant proof of a random feature method.
Since the sketching methods usually works better than the random feature methods (see the examples in \cite{yang2012nystrom}),
the users are advised to apply the sketching methods introduced in Chapter~\ref{sec:spsd}.
Besides the two kinds of low-rank approximation approaches,
the stochastic optimization approach \cite{dai2014scalable} also demonstrates very high scalability.

%%%%%%%%%%%%%%%%%%%%%%%%%%%%%%%%%%%%%%%%%%%%%%%%%%%%%%%%%%%%%%%%%%%%%%%%%%%%%%
%%%%%%%%%%%%%%%%%%%%%%%%%%%%%%%%%%%%%%%%%%%%%%%%%%%%%%%%%%%%%%%%%%%%%%%%%%%%%%
%%%%%%%%%%%%%%%%%%%%%%%%%%%%%%%%%%%%%%%%%%%%%%%%%%%%%%%%%%%%%%%%%%%%%%%%%%%%%%
%%%%%%%%%%%%%%%%%%%%%%%%%%%%%%%%%%%%%%%%%%%%%%%%%%%%%%%%%%%%%%%%%%%%%%%%%%%%%%

\addcontentsline{toc}{chapter}{\bibname}%
\markboth{\bibname}{\bibname}
\bibliography{matrix}

\begin{thebibliography}{36}
\providecommand{\natexlab}[1]{#1}
\providecommand{\url}[1]{\texttt{#1}}
\expandafter\ifx\csname urlstyle\endcsname\relax
  \providecommand{\doi}[1]{doi: #1}\else
  \providecommand{\doi}{doi: \begingroup \urlstyle{rm}\Url}\fi

\bibitem[Ben-Israel and Greville(2003)]{adi2003inverse}
Adi Ben-Israel and Thomas~N.E. Greville.
\newblock \emph{Generalized Inverses: Theory and Applications. Second Edition}.
\newblock Springer, 2003.

\bibitem[Boutsidis and Woodruff(2015)]{boutsidis2015communication}
Christos Boutsidis and David~P Woodruff.
\newblock Communication-optimal distributed principal component analysis in the
  column-partition model.
\newblock \emph{arXiv preprint arXiv:1504.06729}, 2015.

\bibitem[Boutsidis et~al.(2014)Boutsidis, Drineas, and
  Magdon-Ismail]{boutsidis2011near}
Christos Boutsidis, Petros Drineas, and Malik Magdon-Ismail.
\newblock Near-optimal column-based matrix reconstruction.
\newblock \emph{SIAM Journal on Computing}, 43\penalty0 (2):\penalty0 687--717,
  2014.

\bibitem[Charikar et~al.(2004)Charikar, Chen, and
  Farach-Colton]{charikar2004finding}
Moses Charikar, Kevin Chen, and Martin Farach-Colton.
\newblock Finding frequent items in data streams.
\newblock \emph{Theoretical Computer Science}, 312\penalty0 (1):\penalty0
  3--15, 2004.

\bibitem[Clarkson(2005)]{clarkson2005subgradient}
Kenneth~L Clarkson.
\newblock Subgradient and sampling algorithms for l1 regression.
\newblock In \emph{Proceedings of the sixteenth annual ACM-SIAM symposium on
  Discrete algorithms}, pages 257--266. Society for Industrial and Applied
  Mathematics, 2005.

\bibitem[Clarkson and Woodruff(2013)]{clarkson2013low}
Kenneth~L. Clarkson and David~P. Woodruff.
\newblock Low rank approximation and regression in input sparsity time.
\newblock In \emph{Annual ACM Symposium on theory of computing (STOC)}. ACM,
  2013.

\bibitem[Clarkson et~al.(2013)Clarkson, Drineas, Magdon-Ismail, Mahoney, Meng,
  and Woodruff]{clarkson2013fast}
Kenneth~L Clarkson, Petros Drineas, Malik Magdon-Ismail, Michael~W Mahoney,
  Xiangrui Meng, and David~P Woodruff.
\newblock The fast cauchy transform and faster robust linear regression.
\newblock In \emph{Proceedings of the Twenty-Fourth Annual ACM-SIAM Symposium
  on Discrete Algorithms}, pages 466--477. SIAM, 2013.

\bibitem[Cohen and Peng(2014)]{cohen2014lp}
Michael~B Cohen and Richard Peng.
\newblock $\ell_p$ row sampling by lewis weights.
\newblock \emph{arXiv preprint arXiv:1412.0588}, 2014.

\bibitem[Dai et~al.(2014)Dai, Xie, He, Liang, Raj, Balcan, and
  Song]{dai2014scalable}
Bo~Dai, Bo~Xie, Niao He, Yingyu Liang, Anant Raj, Maria-Florina~F Balcan, and
  Le~Song.
\newblock Scalable kernel methods via doubly stochastic gradients.
\newblock In \emph{Advances in Neural Information Processing Systems (NIPS)}.
  2014.

\bibitem[Feldman et~al.(2013)Feldman, Schmidt, and Sohler]{feldman2013turning}
Dan Feldman, Melanie Schmidt, and Christian Sohler.
\newblock Turning big data into tiny data: Constant-size coresets for k-means,
  pca and projective clustering.
\newblock In \emph{Proceedings of the Twenty-Fourth Annual ACM-SIAM Symposium
  on Discrete Algorithms}, pages 1434--1453. SIAM, 2013.

\bibitem[Fowlkes et~al.(2004)Fowlkes, Belongie, Chung, and
  Malik]{fowlkes2004spectral}
Charless Fowlkes, Serge Belongie, Fan Chung, and Jitendra Malik.
\newblock Spectral grouping using the {N}ystr\"{o}m method.
\newblock \emph{IEEE Transactions on Pattern Analysis and Machine
  Intelligence}, 26\penalty0 (2):\penalty0 214--225, 2004.

\bibitem[Gittens(2011)]{gittens2011spectral}
Alex Gittens.
\newblock The spectral norm error of the naive {N}ystr\"om extension.
\newblock \emph{arXiv preprint arXiv:1110.5305}, 2011.

\bibitem[Gittens and Mahoney(2013)]{gittens2013revisiting}
Alex Gittens and Michael~W. Mahoney.
\newblock Revisiting the nystr{\"o}m method for improved large-scale machine
  learning.
\newblock In \emph{International Conference on Machine Learning (ICML)}, 2013.

\bibitem[Halko et~al.(2011)Halko, Martinsson, and Tropp]{halko2011ramdom}
Nathan Halko, Per-Gunnar Martinsson, and Joel~A. Tropp.
\newblock Finding structure with randomness: Probabilistic algorithms for
  constructing approximate matrix decompositions.
\newblock \emph{{SIAM} Review}, 53\penalty0 (2):\penalty0 217--288, 2011.

\bibitem[Johnson and Lindenstrauss(1984)]{johnson1984extensions}
William~B. Johnson and Joram Lindenstrauss.
\newblock Extensions of {L}ipschitz mappings into a {H}ilbert space.
\newblock \emph{Contemporary mathematics}, 26\penalty0 (189-206), 1984.

\bibitem[Mahoney(2011)]{mahoney2011ramdomized}
Michael~W. Mahoney.
\newblock Randomized algorithms for matrices and data.
\newblock \emph{Foundations and Trends in Machine Learning}, 3\penalty0
  (2):\penalty0 123--224, 2011.

\bibitem[Meng and Mahoney(2013)]{meng2013low}
Xiangrui Meng and Michael~W Mahoney.
\newblock Low-distortion subspace embeddings in input-sparsity time and
  applications to robust linear regression.
\newblock In \emph{Proceedings of the forty-fifth annual {ACM} symposium on
  theory of computing}, pages 91--100. ACM, 2013.

\bibitem[Meng et~al.(2014)Meng, Saunders, and Mahoney]{meng2014lsrn}
Xiangrui Meng, Michael~A Saunders, and Michael~W Mahoney.
\newblock Lsrn: A parallel iterative solver for strongly over-or
  underdetermined systems.
\newblock \emph{SIAM Journal on Scientific Computing}, 36\penalty0
  (2):\penalty0 C95--C118, 2014.

\bibitem[Musco and Musco(2015)]{musco2015stronger}
Cameron Musco and Christopher Musco.
\newblock Stronger approximate singular value decomposition via the block
  {L}anczos and power methods.
\newblock \emph{Advances in Neural Information Processing Systems (NIPS)},
  2015.

\bibitem[Nystr{\"o}m(1930)]{nystrom1930praktische}
Evert~J. Nystr{\"o}m.
\newblock {\"U}ber die praktische aufl{\"o}sung von integralgleichungen mit
  anwendungen auf randwertaufgaben.
\newblock \emph{Acta Mathematica}, 54\penalty0 (1):\penalty0 185--204, 1930.

\bibitem[Pham and Pagh(2013)]{pham2013fast}
Ninh Pham and Rasmus Pagh.
\newblock Fast and scalable polynomial kernels via explicit feature maps.
\newblock In \emph{the 19th ACM SIGKDD international conference on Knowledge
  discovery and data mining (KDD)}, pages 239--247. ACM, 2013.

\bibitem[Rahimi and Recht(2007)]{rahimi2007random}
Ali Rahimi and Benjamin Recht.
\newblock Random features for large-scale kernel machines.
\newblock In \emph{Advances in neural information processing systems (NIPS)},
  pages 1177--1184, 2007.

\bibitem[Saad(2011)]{saad2011numerical}
Yousef Saad.
\newblock Numerical methods for large eigenvalue problems.
\newblock \emph{preparation. Available from: http://www-users. cs. umn.
  edu/saad/books. html}, 2011.

\bibitem[Si et~al.(2014)Si, Hsieh, and Dhillon]{si2014memory}
Si~Si, Cho-Jui Hsieh, and Inderjit Dhillon.
\newblock Memory efficient kernel approximation.
\newblock In \emph{International Conference on Machine Learning (ICML)}, pages
  701--709, 2014.

\bibitem[Stewart(1999)]{stewart1999four}
G.~W. Stewart.
\newblock Four algorithms for the efficient computation of truncated pivoted
  {QR} approximations to a sparse matrix.
\newblock \emph{Numerische Mathematik}, 83\penalty0 (2):\penalty0 313--323,
  1999.

\bibitem[Thorup and Zhang(2012)]{thorup2012tabulation}
Mikkel Thorup and Yin Zhang.
\newblock Tabulation-based 5-independent hashing with applications to linear
  probing and second moment estimation.
\newblock \emph{SIAM J. Comput.}, 41\penalty0 (2):\penalty0 293--331, April
  2012.
\newblock ISSN 0097-5397.

\bibitem[Tropp(2015)]{tropp2015introduction}
Joel~A Tropp.
\newblock An introduction to matrix concentration inequalities.
\newblock \emph{arXiv preprint arXiv:1501.01571}, 2015.

\bibitem[Wang et~al.(2015{\natexlab{a}})Wang, Li, Mahoney, and
  Darve]{wang2015structured}
Ruoxi Wang, Yingzhou Li, Michael~W Mahoney, and Eric Darve.
\newblock Structured block basis factorization for scalable kernel matrix
  evaluation.
\newblock \emph{arXiv preprint arXiv:1505.00398}, 2015{\natexlab{a}}.

\bibitem[Wang and Zhang(2013)]{wang2013improving}
Shusen Wang and Zhihua Zhang.
\newblock Improving {CUR} matrix decomposition and the {N}ystr\"{o}m
  approximation via adaptive sampling.
\newblock \emph{Journal of Machine Learning Research}, 14:\penalty0 2729--2769,
  2013.

\bibitem[Wang et~al.(2014)Wang, Luo, and Zhang]{wang2014modified}
Shusen Wang, Luo Luo, and Zhihua Zhang.
\newblock Spsd matrix approximation via column selection: Theories, algorithms,
  and extensions.
\newblock \emph{CoRR}, abs/1406.5675, 2014.

\bibitem[Wang et~al.(2015{\natexlab{b}})Wang, Zhang, and
  Zhang]{wang2015towards}
Shusen Wang, Zhihua Zhang, and Tong Zhang.
\newblock Towards more efficient symmetric matrix sketching and {CUR} matrix
  decomposition.
\newblock \emph{arXiv preprint arXiv:1503.08395}, 2015{\natexlab{b}}.

\bibitem[Williams and Seeger(2001)]{williams2001using}
Christopher Williams and Matthias Seeger.
\newblock Using the {N}ystr{\"o}m method to speed up kernel machines.
\newblock In \emph{Advances in Neural Information Processing Systems (NIPS)},
  2001.

\bibitem[Wilson and Nickisch(2015)]{wilson2015kernel}
Andrew~Gordon Wilson and Hannes Nickisch.
\newblock Kernel interpolation for scalable structured gaussian processes
  (kiss-gp).
\newblock \emph{arXiv preprint arXiv:1503.01057}, 2015.

\bibitem[Woodruff(2014)]{woodruff2014sketching}
David~P Woodruff.
\newblock Sketching as a tool for numerical linear algebra.
\newblock \emph{arXiv preprint arXiv:1411.4357}, 2014.

\bibitem[Yang et~al.(2012)Yang, Li, Mahdavi, Jin, and Zhou]{yang2012nystrom}
Tianbao Yang, Yu-Feng Li, Mehrdad Mahdavi, Rong Jin, and Zhi-Hua Zhou.
\newblock {Nystr\"om} method vs random fourier features: A theoretical and
  empirical comparison.
\newblock In \emph{Advances in Neural Information Processing Systems (NIPS)},
  2012.

\bibitem[Zhang and Kwok(2010)]{zhang2010clustered}
Kai Zhang and James~T. Kwok.
\newblock Clustered {Nystr{\"o}m} method for large scale manifold learning and
  dimension reduction.
\newblock \emph{IEEE Transactions on Neural Networks}, 21\penalty0
  (10):\penalty0 1576--1587, 2010.

\end{thebibliography}

\end{document}